# Valorization of byproducts of hemp multipurpose crops: short non-aligned bast fibers as a source of nanocellulose

Sara Dalle Vacche[a,*], Vijayaletchumy Karunakaran[a,b], Alessia Patrucco[c], Marina Zoccola[c], Loreleï Douard[d], Silvia Maria Ronchetti[a], Marta Gallo[a], Aigoul Schreier[e], Yves Leterrier[e], Julien Bras[d], Davide Beneventi[d], Roberta Bongiovanni[a]

[a] Politecnico di Torino, Department of Applied Science and Technology, Corso Duca degli Abruzzi 24, 10129 Turin, Italy

[b] Universiti Kuala Lumpur Malaysian Institute of Chemical and Bioengineering Technology (UniKL MICET), Lot 1988 Kawasan Perindustrian Bandar Vendor, Taboh Naning, 78000 Alor Gajah, Melaka, Malaysia

[c] Consiglio Nazionale delle Ricerche, Istituto di Sistemi e Tecnologie Industriali Intelligenti per il Manifatturiero Avanzato, Biella, Italy

[d] Université Grenoble Alpes, CNRS, Grenoble INP, LGP2, Grenoble, France

[e] Laboratory for Processing of Advanced Composites (LPAC), Ecole Polytechnique Fédérale de Lausanne (EPFL), CH-1015 Lausanne, Switzerland

[*]Corresponding author

Email addresses: sara.dallevacche@polito.it (Sara Dalle Vacche), vijayaletchumy13@s.unikl.edu.my (Vijayaletchumy Karunakaran), alessia.patrucco@stiima.cnr.it (Alessia Patrucco), marina.zoccola@stiima.cnr.it (Marina Zoccola), Lorelei.Douard@lgp2.grenoble-inp.fr (Loreleï Douard), davide.beneventi@pagora.grenoble-inp.fr (Davide Beneventi), marta.gallo@polito.it (Marta Gallo), silvia.ronchetti@polito.it (Silvia Ronchetti) julien.bras@grenoble-inp.fr (Julien Bras),

aigoul.schreier@epfl.ch (Aigoul Schreier), yves.leterrier@epfl.ch (Yves Leterrier),

roberta.bongiovanni@polito.it (Roberta Bongiovanni)

**Abstract**

Nanocellulose was extracted from short bast fibers, from hemp (*Cannabis Sativa* L.) plants harvested at seed maturity, non-retted, mechanically decorticated in a defibering apparatus giving non-aligned fibers. A chemical pretreatment with NaOH and HCl allowed the removal of most of the non-cellulosic components of the fibers. No bleaching was performed. The chemically pretreated fibers were then refined in a beater and treated with a cellulase enzyme, before the mechanical defibrillation in an ultrafine friction grinder. The fibers were characterized by microscopy, infrared spectroscopy, thermogravimetric analysis and X-ray diffraction after each step of the process, to understand the evolution of their morphology and composition. The obtained nanocellulose suspension was composed by rod-like fibrils with widths of 5-12 nm, stacks of nanofibrils with widths of 20-200 nm, and some larger fibers. The crystallinity index was found to increase from 74% for the raw fibers to 80% for the nanocellulose. The nanocellulose retained a yellowish color indicating the presence of some residual lignin. Properties of nanopaper prepared with the hemp nanocellulose were similar to those of nanopapers prepared with wood pulp derived rod-like nanofibrils.

## 1. Introduction

The varieties of *Cannabis sativa* L. with δ-tetrahydrocannabinol (THC) levels below the legal limit are commonly referred to as industrial hemp, or simply hemp. Hemp is an annual crop, suitable for the European climates. It is sustainable as it requires a low water and fertilizer input, and has no need for herbicides and insect repellants; in addition, it improves soil structure, acting as an excellent break crop in crop rotations, potentially enhancing the yield of other food crops, and can adapt to marginal or contaminated land (Amaducci and Gusovius, 2010; Angelini et al., 2016). For these reasons, after a period that saw the decline of hemp cultures due to anti-drug legislation, Europe sees a growing interest in industrial hemp. The Italian region of Piedmont has a long history in hemp cultivation, particularly of the local dioecious Carmagnola variety, a giant variety with a very tall stem and high quality fibers (McPartland, 2020). Once grown for textile applications, it is now primarily grown for the production of seeds, and for the inflorescences and extraction biomass rich of cannabidiol. The stems remain thus as a byproduct, that has to be exploited to make hemp cultivation truly profitable for farmers (Amaducci et al., 2015).

The inner woody core of the stem is composed of short and more lignified fibers, the so-called shives, used mainly as animal bedding, garden mulch or construction material (Guerriero et al., 2016). The core is surrounded by bast fibers, which are of two types: the outer ones, called primary fibers, are thicker, longer and richer in cellulose, while the ones closer to the inner core, the secondary fibers, are thinner, shorter and contain a higher amount of lignin (Crônier et al., 2005). The primary bast fibers from plants harvested at full flowering are suitable for textile applications or high-performance composites: to this aim after retting they are separated from the shives, with techniques that allow collecting long and aligned fiber bundles (Musio et al., 2018; Placet, 2009). When the cultivation of hemp is multipurpose, the plant is grown to a higher degree of maturity, leading to a higher lignification of the fibers and to an increased content of secondary bast fibers; the stems are then typically processed with so-called “disordered” decortication methods. The fibers recovered in this way are non-aligned, shorter and of lower quality (Amaducci et al., 2005; Amaducci and Gusovius, 2010), thus unsuitable for many industries such as textiles, high-performance composites or quality paper (Musio et al., 2018; Tang et al., 2016; Westerhuis et al., 2019).

Bast fibers are multicellular and have a hierarchical structure. They are bundles of single cells called individual fibers, cemented together by the middle lamella, a layer composed primarily of lignin, pectin, and small amounts of proteins (Zamil and Geitmann, 2017). Each individual fiber has a hollow central part called lumen, surrounded by a thick secondary cell wall and a thinner primary cell wall. The cell walls are made of cellulose having a semi-crystalline fibrillary structure, the so-called microfibrils, embedded in an amorphous matrix of hemicellulose, pectin and lignin (Ansell and Mwaikambo, 2009; Garside and Wyeth, 2003; Melelli et al., 2020; Wang et al., 2019). Traditionally, crystalline and amorphous cellulose regions are thought to alternate along the microfibrils, while recent studies suggest that amorphous cellulose chains may be located preferentially at the surface of the microfibrils, and the crystalline cellulose chains in the interior (Garvey et al., 2005). Finally, small amounts of bound water, waxes, pigments and traces of inorganic compounds are also present in the fibers.

Hemp fibers, together with many other fibers from annual plants, have been proposed as a lignocellulosic biomass for the production of nanocellulose, both nanofibrillated and nanocrystalline. Cellulose in hemp fibers has been reported to range from about 55% to 80% depending on the variety and maturity of the plant, and the type of fiber, i.e. primary or secondary bast fibers, or shives (Crônier et al., 2005; Korte and Staiger, 2008; Manaia et al., 2019; Marrot et al., 2013); the lignin content is lower than in wood, particularly in the bast fibers, allowing for milder fiber delignification and purification processes (Mondragon et al., 2014; Nechyporchuk et al., 2016). Some attempts have therefore been made to use hemp fibers for the production of nanocellulose through chemical, enzymatic, and/or mechanical defibrillation methods. Most of the processes either use bleached pulps or include a bleaching step, or involve an oxidative treatment, such as TEMPO (2,2,6,6-tetramethylpiperidine-1-oxyl) mediated oxidation (Abraham et al., 2016; Dai et al., 2013; Luzi et al., 2014; Pacaphol and Aht-Ong, 2017; Puangsin et al., 2013b, 2013a; Wang et al., 2007).

Bleaching performed with chlorine or (hypo)chlorite, the latter being also used for TEMPO oxidation, is a common procedure. However, it results in the release of organic halides and other toxic compounds (Singh and Chandra, 2019). Elemental chlorine free (ECF) pulp bleaching processes, using chlorine dioxide or alkali $H_2O_2$, are more environmentally acceptable (Santos and Hart, 2013), despite concerns related to health and explosion hazards (Hart and Rudie, 2010; Singh and Chandra, 2019). Also the sustainability of the current $H_2O_2$ production process is under scrutiny (Campos-Martin et al., 2006). Greener approaches, avoiding both bleaching and oxidation, have been recently proposed for the preparation of nanocellulose from hemp fibers and shives: main examples are steam explosion followed by ball milling and sonication (Šutka et al., 2013), dual asymmetrical centrifugation (Agate et al., 2020), or enzymatic treatment with endoglucanase followed by sonication, combined with a pretreatment with NaOH, ultrasounds and microwaves (Xu et al., 2013).

In this work we report the production of nanocellulose from non-aligned hemp bast fibers from multipurpose crop, to provide added value applications for these abundant and cheap agricultural residues. We used a chemical pretreatment with alkali and acid, adapted from Wang et al. (Wang et al., 2007), to remove non cellulosic compounds, followed by an enzymatic pretreatment and finally by mechanical defibrillation with an ultrafine friction grinder. As the raw fibers contained only a very small amount of lignin, in the effort of providing a more environmentally friendly process we avoided bleaching and other oxidative treatments. The fibers were analyzed after the pre-treatment steps; to evaluate the properties of the obtained nanocellulose, and assess its performance for possible applications, nanopaper was prepared and characterized in this work.

## 2. Materials and methods

### *2.1 Materials*

Non-aligned bast fibers from hemp (*Cannabis sativa* L.) plants of Carmagnola variety, harvested at seed maturity, non-retted, and mechanically decorticated in a “disordered line”, were obtained from Assocanapa (Carmagnola, TO, Italy). The fibers were cleaned and disentangled with a Mixicomber machine (O. M. Alvaro Mason & C S.N.C., Italy) prior to use.

The chemicals used for the pretreatment were sodium hydroxide pellets (Carlo Erba Reagents S.A.S., Val de Reuil, France) and hydrochloric acid, 37% (Sigma-Aldrich, Inc., St. Louis, MO, US); ultrapure water spilled from a Millipore Direct-Q 3 UV system (Merck KGaA, Darmstadt, Germany) was used to dilute them at the desired concentrations. For the enzymatic treatment a buffer was prepared using acetic acid ≥ 99.7% and sodium acetate trihydrate ≥ 99.0% (Sigma-Aldrich, Inc., St. Louis, MO, US); the enzyme was FiberCare® R cellulase 4890 ECU (Novozymes, Copenhagen, Denmark).

*2.2 Production of nanocellulose from hemp fibers*

The disentangled hemp fibers (raw fibers) were chopped using a kitchen grinder, and then underwent a number of chemical and mechanical pretreatments before the final mechanical fibrillation step, as detailed in what follows; they were never dried during the procedure. After each pre-treatment step, samples of fibers were set aside for analysis.

The chopped hemp fibers were soaked in a 10 wt% aqueous sodium hydroxide (NaOH) solution at room temperature for two hours, with mild agitation, then in a 1M aqueous solution of hydrochloric acid (HCl) for one hour at 80 °C, and finally, in a 2 wt% aqueous NaOH solution for one hour at 80 °C. The ratio of liquid to fibers was equal to 1 l of each solution per 20 g of dry chopped hemp fibers. Between each step the fibers were rinsed with water, with the aid of a 120-mesh sieve, until the pH was in the 6 to 8 range, and were drained from excess water, but not allowed to dry.

The chemically pre-treated fibers were diluted to a 10 wt% concentration in demineralized water and processed with an immersion kitchen blender to obtain a finer pulp. The pulp was then further refined in a PFI mill-type laboratory beater (Metrotec SA) at 2500 rpm, prior to the treatment with the FiberCare® cellulase enzyme. The concentration of the enzyme was 300 ECU per gram of dry fibers. The refined fibers, placed in a reactor equipped with an overhead stirrer and a heated water bath, were further diluted to a 2 wt% concentration using a pH 5 acetic acid / sodium acetate trihydrate buffer solution. The temperature was raised to 50 °C, then the enzymes were added: the fibers were left to stir for 2 hours, before raising the temperature to 80 °C for 10 min to deactivate the enzymes. The pulp was then vacuum filtered using a precision woven screening polyamide 6.6 fabric with 1 µm mesh opening (NITEX, Sefar AG), and re-dispersed in demineralized water. The filtration procedure was repeated three times in order to reach a neutral pH.

Finally, to obtain the nanocellulose, demineralized water was added to the enzymatically treated pulp, to obtain a solids content of approximately 1.4 g/l. The pulp was defibrillated in a Supermasscolloider ultra-fine friction grinder (Model MKZA6-2, disk model MKG-C 80, Masuko Sangyo Co., Ltd.). The clearance gauge was set initially at 0 position, and the suspension was passed 8 times through the grinder in order to homogenize it. Then the suspension was passed through the grinder 4 times with the gauge set at -5, 8 times with the gauge set at -10, 1 time with the gauge at -12, and finally 2 times with the gauge at -15. A small sample of the nanocellulose aqueous suspension was dried overnight at 100 °C in order to verify the actual concentration, which was found to be 1.3 wt%.

*2.3 Production of nanopaper*

Nanopaper handsheets were prepared using the 1.3 wt% nanocellulose aqueous suspension, as obtained after the ultrafine friction grinding, by filtering through a Rapid-Köthen standard sheet former (Frank-PTI, Germany) equipped with a precision woven screening polyamide 6.6 fabric with 1 µm mesh opening (NITEX, Sefar AG). The sheets were then dried in a dryer at 90 °C for 20 minutes, between two layers of the polyamide 6.6 fabric. The final thickness was 55±1 µm.

*2.4 Characterization methods*

For measuring the insoluble lignin an acid treatment was applied to the fibers to completely hydrolyze cellulose and hemicellulose. First the fibers were treated with a 72% $H_2SO_4$ solution for 2h at room temperature, and then the acid solution was diluted to 3% and the temperature was raised to the boiling point for 4 h. Then the suspension was left to decant overnight, and then was filtered to recover the insoluble lignin, which was weighed after drying overnight. For the determining the soluble lignin content, 50 ml of the filtrate of the hydrolysis was collected, and the absorbance at 205 nm was measured with a UV-Vis spectrophotometer, using a 3% $H_2SO_4$ solution as blank.

The yield, expressed as a percentage, after the chemical pretreatment and after the entire process was calculated as yield (%) = 100 · $M_i/M_{chop}$, where $M_{chop}$ was the mass of chopped fibers, and $M_i$ was the mass of fibers after the selected step, i.e. after the chemical pretreatment, and after the ultrafine friction grinding.

Samples of the hemp fibers taken after each pre-treatment step were examined in reflection mode with a BX53M optical microscope (Olympus Italia S.R.L.) and in transmission mode with an Axio Imager M1 optical microscope equipped with an AxioCam MRc 5 digital camera (Carl Zeiss). For the observation the fibers were dispersed at low concentration in distilled water, a small amount was poured on a glass slide, and covered with a glass lid. Images were taken at various magnifications. Samples were also observed with a field emission scanning electron microscope (Supra 40 FE-SEM, Carl Zeiss). The pretreated fibers were dried in an oven at 60 °C and spread on carbon tape. The nanocellulose suspension was further diluted with addition of distilled water, and a droplet was poured on a transmission electron microscope (TEM) grid and left to evaporate at room temperature. The specimens were coated with Cr.

In order to obtain transmission electron microscope (TEM) images of the nanosized particles, a small sample of the nanocellulose suspension was diluted and decanted to separate the larger fibers. Droplets of the decanted nanocellulose suspension were deposited onto glow-discharged carbon-coated TEM grids. After a few minutes, the liquid in excess was blotted with filter paper and the preparation was negatively stained with 2 wt% uranyl acetate. The stain in excess was blotted and the specimen allowed to dry. The specimens were observed with a JEOL JEM-2100-Plus microscope operating at 200 kV and images were recorded with a Gatan Rio 16 digital camera.

The dimensions (length and width) of the fibers after each pre-treatment step was assessed from the optical microscope, FESEM and TEM images by means of the software ImageJ (release 1.52s, Wayne Rasband, National Institutes of Health, USA)

Fourier Transform Infrared (FT-IR) analysis was performed with a Nicolet iS50 spectrometer (Thermo Fisher Scientific Inc.), equipped with a Smart iTX-Diamond Attenuated Total Reflectance (ATR) accessory, in the 4000-550 $cm^{-1}$ range with 32 scans per spectrum and a 4 cm-1 resolution. To obtain a flat surface for analysis, the samples were poured in a glass Petri dish and dried at 60 °C until constant weight was reached. These samples were then directly put on the FTIR diamond for analysis.

Thermogravimetric analysis (TGA) was performed from 25 °C to 800 °C with a heating rate of 20 °C $min^{-1}$, under a 60 ml $min^{-1}$ $N_2$ flux, to prevent thermo-oxidative processes (TGA/SDTA 851e, Mettler Toledo). Prior to analysis all samples were dried in an oven at 100 °C for 1 hour.

For differential scanning calorimetry (DSC), for each sample two heating cycles were performed at temperatures between -70 °C and 250 °C at a heating rate of 10 °C $min^{-1}$ under $N_2$ flow (DSC 1 STARe System, Mettler Toledo).

X-ray diffraction characterization was performed on samples of dry fibers and nanocellulose using a Panalytical X'Pert PRO (Cu Kα radiation) diffractometer, with a PIXcel detector, a solid-state detector with rapid readout time and high dynamic range. Data collection has been performed at 40 kV and 40 mA, between 10° and 40° 2θ, with a step of 0.02° 2θ and a wavelength of 1.54187 Å. To allow rapid comparison of fiber samples, the crystallinity index (CI) was calculated according to Segal equation (Segal et al., 2016):

$$CI = 100 \times (I_{200} - I_{am})/I_{200} \qquad (1)$$

where $I_{200}$ is the height of the 200 peak and $I_{am}$ the height of the minimum diffraction intensity between the 200 and the 110 peaks. This method is useful for comparing the relative differences between samples (Park et al., 2010). Crystallites size was determined according to Scherrer equation (Scherrer, 1918) by comparing the profile width of a standard profile with the sample profile. The Scherrer equation relates the width of a powder diffraction peak to the average dimensions of crystallites in a polycrystalline powder:

$$D = K\lambda / \beta_{(2\theta)hkl} \cos\theta \qquad (2)$$

where $\beta$ is the crystallite size contribution to the peak width at half maximum (FWHM) in radians, $K$ (shape factor) is a constant which value is close to unity, and $D$ is the average thickness of the crystal in a direction normal to the diffracting plane *hkl*. Profile fits were performed using X'Pert High Score Plus, using Pseudo-Voigt peak function with $K\alpha_1$ and $K\alpha_2$ fitting on a background stripped pattern. The sample-induced peak broadening $\beta$ was determined by subtracting the instrumental peak width from the measured peak width. The instrumental broadening was determined using $LaB_6$ powder (NIST SRM®660a, size of crystallites in the 2 µm - 5 µm range).

In order to assess the compatibility of the hemp nanocellulose with different solvents, demineralized water, ethanol, or acetone were added to the nanocellulose suspension, to obtain a 0.25 wt.% nanocellulose concentration. The suspensions were then homogenized with a T10 Ultra Turrax and put into closed vials; their settling was observed during 12 days.

Dynamic mechanical analysis (DMA) was performed in tensile mode on rectangular specimens of nanopaper handsheet. The temperature was increased from -100 °C to 180 °C at a 3 °C $min^{-1}$ rate, the frequency was set at 1 Hz, and the strain was set at 0.01% (TTDMA, Triton Technology Ltd., UK). The specimens had a length of 10 mm between the clamps, and a width of 6 to 8 mm.

Tensile tests were performed on rectangular specimens cut from the handsheet, using a Dual Leadscrew 5KN tensile testing stage (Deben UK Ltd., Suffolk, UK), with a speed of 1.0 mm/min.

To measure the Oxygen Transmission Rate (OTR) and the Water Vapor Transmission Rate (WVTR), an oxygen permeation analyzer Systech 8001, and a water vapour permeation analyzer Systech 7001 (both by Systech Illinois), were used, respectively. In the two parallel chambers of each instrument, specimens of the nanopaper were mounted using a steel mask with a circular opening of 5 $cm^2$. During the conditioning cycle dry nitrogen was used to purge the chambers. Then for the measurements a flow of either pure oxygen gas (for OTR) or wet nitrogen gas (for WVTR) was fluxed to one side of the film. The OTR measurement was performed at 23 °C and 50% RH, while the WVTR measurement was performed at 38 °C and 50% RH. The results were collected when the oxygen or water vapor transmission rates reached a steady-state. In the text that follows all the reported OTR values are expressed in $cm^3\ m^{-2}\ day^{-1}\ bar^{-1}$ and the oxygen permeability ($P_{O2}$) values in $cm^3\ mm\ m^{-2}\ day^{-1}\ bar^{-1}$, the reported WVTR values are expressed in $g\ m^{-2}\ day^{-1}$ and the water vapor permeability ($P_{H2O}$) values in $g\ mm\ m^{-2}\ day^{-1}\ atm^{-1}$: these units will be omitted to improve readability.

## 3. Results and discussion

The composition of  Carmagnola hemp bast fibers collected after dry-mechanical treatment of the stalks was reported to be approximately 61-69% cellulose, 14-17% hemicellulose, and  3-5% lignin (Cappelletto et al., 2001). Similarly, the raw fibers used in this work contained a low amount of lignin, i.e. 3.9% insoluble lignin and 0.7% soluble lignin.

The process for obtaining the nanocellulose, detailed in the experimental part, is summarized in Figure 1: the fibers were chopped, then underwent a number of pretreatments with alkali, acid and a cellulase enzyme, before the final mechanical defibrillation step. Characterization was performed on the hemp fibers after each pre-treatment step, in order to elucidate the effect of each pre-treatment on the fiber structure, and on the final nanocellulose. The yield after the chemical pretreatment was about 60%, which accounts for the removal of non-cellulosic compounds, losses of smaller fibers through the sieve during the rinsing steps, and the removal of adsorbed water in the raw fibers upon drying. This result is compatible with the cellulosic content reported in literature for Carmagnola hemp fibers. At the end of the treatment, the nanocellulose suspension recovered after the mechanical defibrillation accounted for 97% of the weight of the chemically treated fibers, therefore resulting in a global yield of 57%. The nanocellulose suspension, as obtained after mechanical defibrillation, was used to prepare nanopaper, which was then characterized.

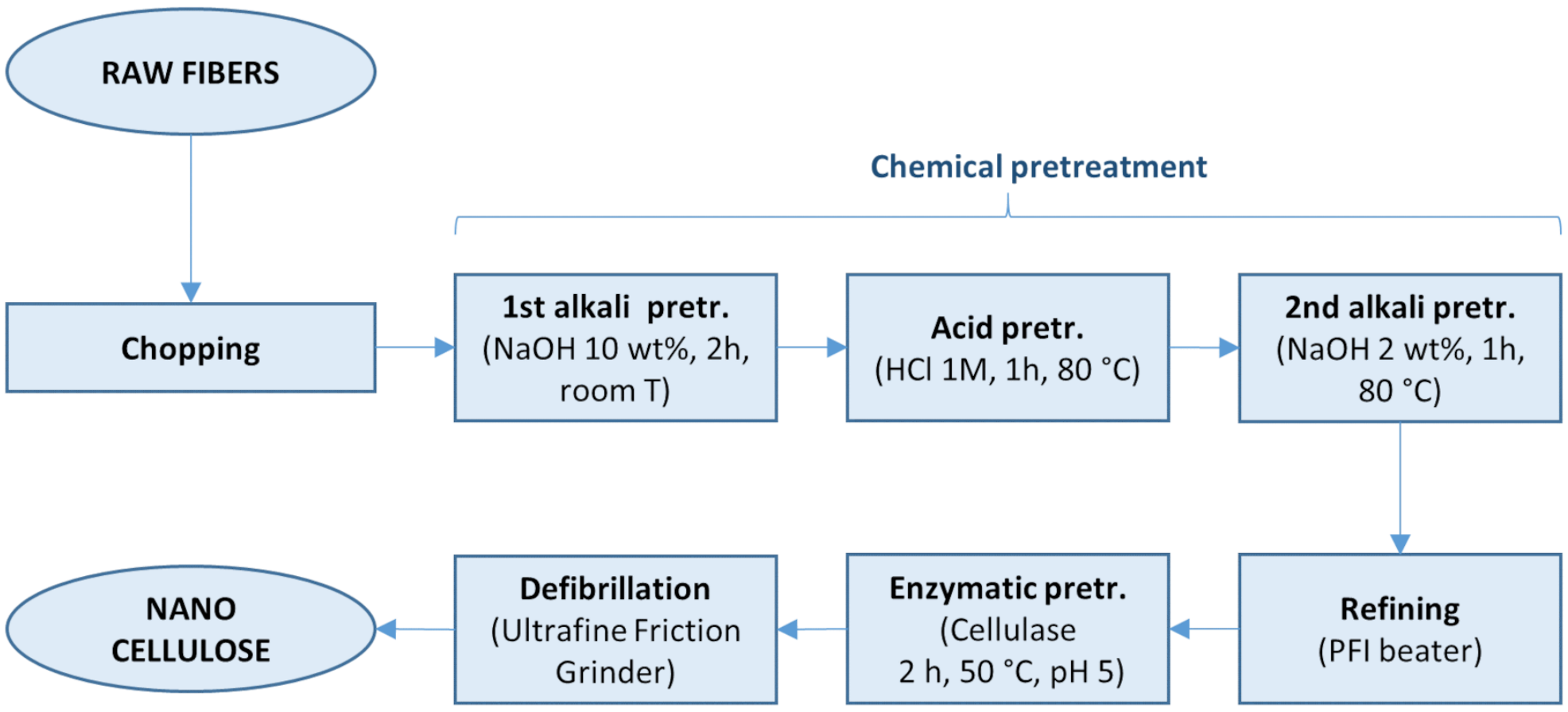

Figure 1. Scheme of the process for obtaining hemp nanocellulose

### *3.1 Evolution of the dimensions and morphology of the fibers during the pretreatment*

A photo of the hemp fibers, chopped and after each step of the chemical pretreatment, showing the color change of the fibers, is shown in Figure 2.

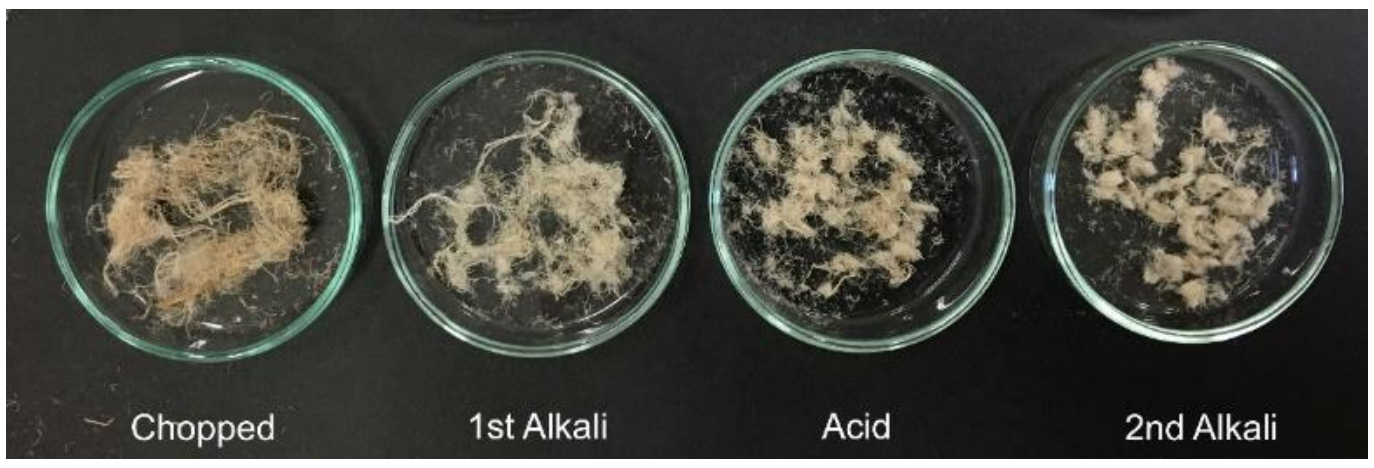

Figure 2. Fibers samples showing the color change after different steps of the chemical pretreatment

After the first alkali pretreatment, with a 10 wt% NaOH solution for 2 hours at room temperature, the fibers had a lighter color than the chopped ones, due to the removal of the non-cellulosic compounds of the cell wall, i.e. hemicellulose and lignin. Accordingly, the alkali solution recovered after the treatment had a dark yellow color. After the acid and the second alkali pretreatment, with 1M HCl and with 2 wt% NaOH, respectively, both performed for 1 hour at 80 °C, the fibers still retained a yellowish color. As the corresponding solutions recovered at the end of these treatments had also a pale-yellow color, it is inferred that some hemicellulose and lignin that were protected inside the fibers were however further removed during these steps.

The evolution of the dimensions of the fibers during the chemical pretreatment is shown in Figure 3, and Figure 4 shows FESEM images of the surface of the fibers after each pretreatment step. Additional optical microscopy and FESEM images can be found in the open access Zenodo repository (Dalle Vacche et al., 2021a).

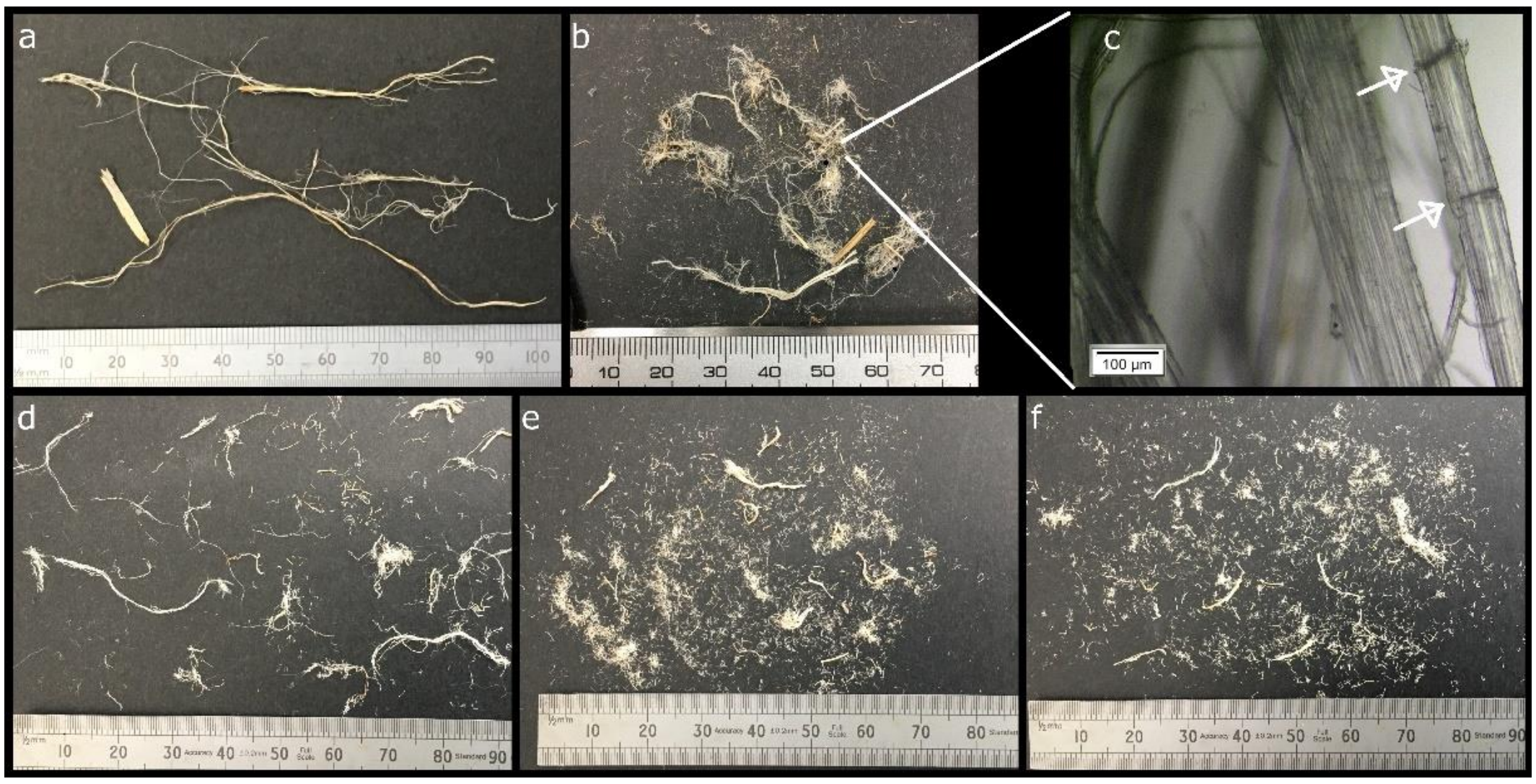

Figure 3. Images showing the evolution of the dimensions of the fibers: (a) raw fibers; (b) chopped fibers (c) optical micrograph of chopped fibers in which dislocations appear as transversal dark lines, indicated by the arrows; (d) fibers after the 1st alkali treatment (e) fibers after the acid treatment (f) fibers after the 2nd alkali treatment.

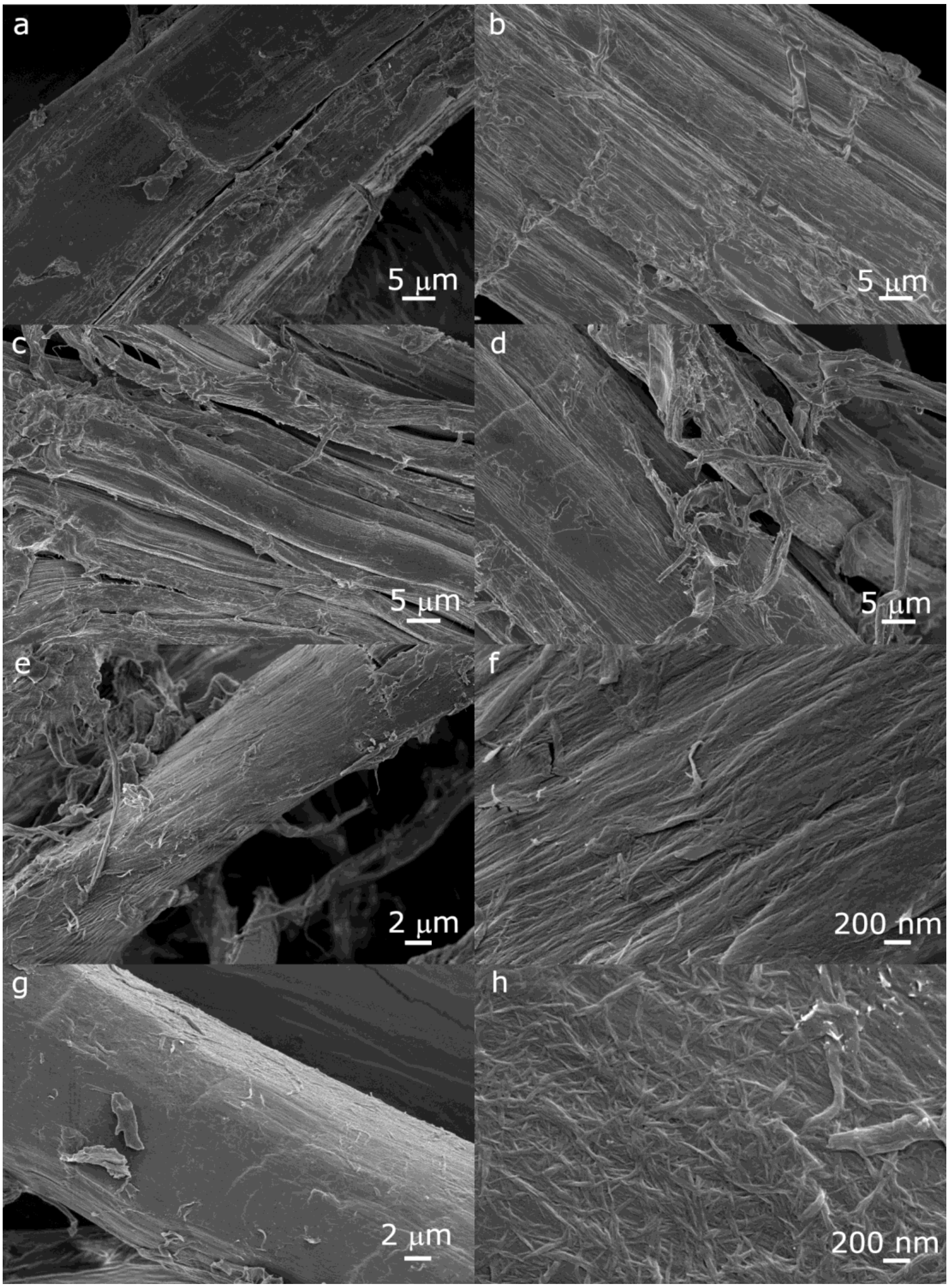

Figure 4. FESEM images of the fibers after each pretreatment step, showing the appearance of their surface: (a) chopped fibers (b) after the 1st alkali treatment (c) after the acid treatment (d) after the 2nd alkali treatment, (e, f) after refining (g, h) after the enzymatic treatment. Images (f, h) show that the microfibrils start appearing at the surface of the fibers after refining and enzymatic pretreatment.

The raw hemp fiber sample was composed of primary and secondary bast fibers having a wide range of dimensions, with widths of few micrometers to several hundred micrometers and lengths up to 150 mm, and a small number of fragments of shives, with widths of up to 2 mm and lengths up to 10 – 20 mm.

After chopping, the maximum fiber length was reduced to about 40 mm, with a large amount of shorter fibers, so that the median length was about 5 mm; dislocations (defects), appearing as transversal dark lines in the optical micrographs, could be observed. The mechanical action of the mixer blades not only cut the fibers, but also favored the splitting of the bundles of individual fibers. The fiber bundles had sections of irregular shapes, while the individual fibers had more round sections. Fiber widths were mostly in the 1-500 µm range, with large part of the fibers having widths below 80 µm. Some fragments of shives having widths up to about 1 mm were still visible. The surface of the fiber bundles was rough as the middle lamella still surrounded the individual fibers; few thin fibers with micron to submicron dimensions departed from the surface of the fiber bundles (Figure 4a).

After the first alkali pretreatment the fibers still had lengths up to 40 mm, although, with respect to the chopped fibers, the number of shorter fibers increased, with the median value for the length decreasing to 2 mm. The widths of the fibers were similar to those of the chopped fibers. Due to the removal of the middle lamella, the individual fibers composing the bundles became clearly visible, but did not much further separate (Figure 4b), as even high concentrations of NaOH were reported not to be able to completely remove all the lignin and pectin cementing the bundles (Mwaikambo and Ansell, 2006). The estimated widths of the individual fibers, 3 µm to 20 µm, were consistent with those reported in literature (Mwaikambo and Ansell, 2006; Tatyana et al., 2017).

The acid pretreatment overall reduced the length of the fibers, likely due to acid hydrolysis that is known to weaken the fibers at dislocations (Thygesen, 2008). The observed lengths were in the 0.1 mm to 16 mm range, with a median length of 0.9 mm. This range of lengths is in line with that reported for hemp individual fibers (Tatyana et al., 2017). The treatment favored the separation of the individual fibers (Figure 4c), as also indicated in other works (Abraham et al., 2016). As a result, the amount of finer fibers increased, with the median value for fiber width decreasing to about 16 μm, although the upper limit for the observed fiber widths did not change, as not all the bundles were disrupted.

The second alkali pretreatment further loosened the individual fibers, removing non-cellulosic compounds; some thin ribbon-like filaments, probably parts of damaged individual fibers, could be seen protruding from their surfaces (Figure 4d). A slightly higher amount of shorter fibers could be observed, while the observed widths did not substantially change with respect to the previous step.

Refining greatly reduced the length and width of the fibers. The fiber length was in the 400 μm to 10 μm range, with a median value of about 70 μm. The majority of the fibers had widths ranging from submicron to about 20 μm, and the median value was close to 3 μm, and on the surface of the individual fibers, a fibrous structure of nanometric fibrils appeared (Figure 4e,f).

The enzymatic pretreatment did not further change significantly the fiber dimensions; nevertheless, the surface of the fibers was “cleaner” and a structure of rod-like fibrils of nanometric dimensions became well visible (Figure 4g,h).

To summarize, the dimensions of the hemp fibers after each pre-treatment step are compared in the box plots of Figure 5. One can conclude that a reduction of the length was obtained gradually throughout the pre-treatments, with an abrupt decrease after refining. On the other hand, the width of the fibers did not decrease sensibly with the chemical pretreatments, while a significant decrease was obtained with mechanical refining, suggesting that mechanical energy input is needed for separating the individual fibers. The pretreatment with enzymes did not significantly further reduce the fiber lengths and widths.

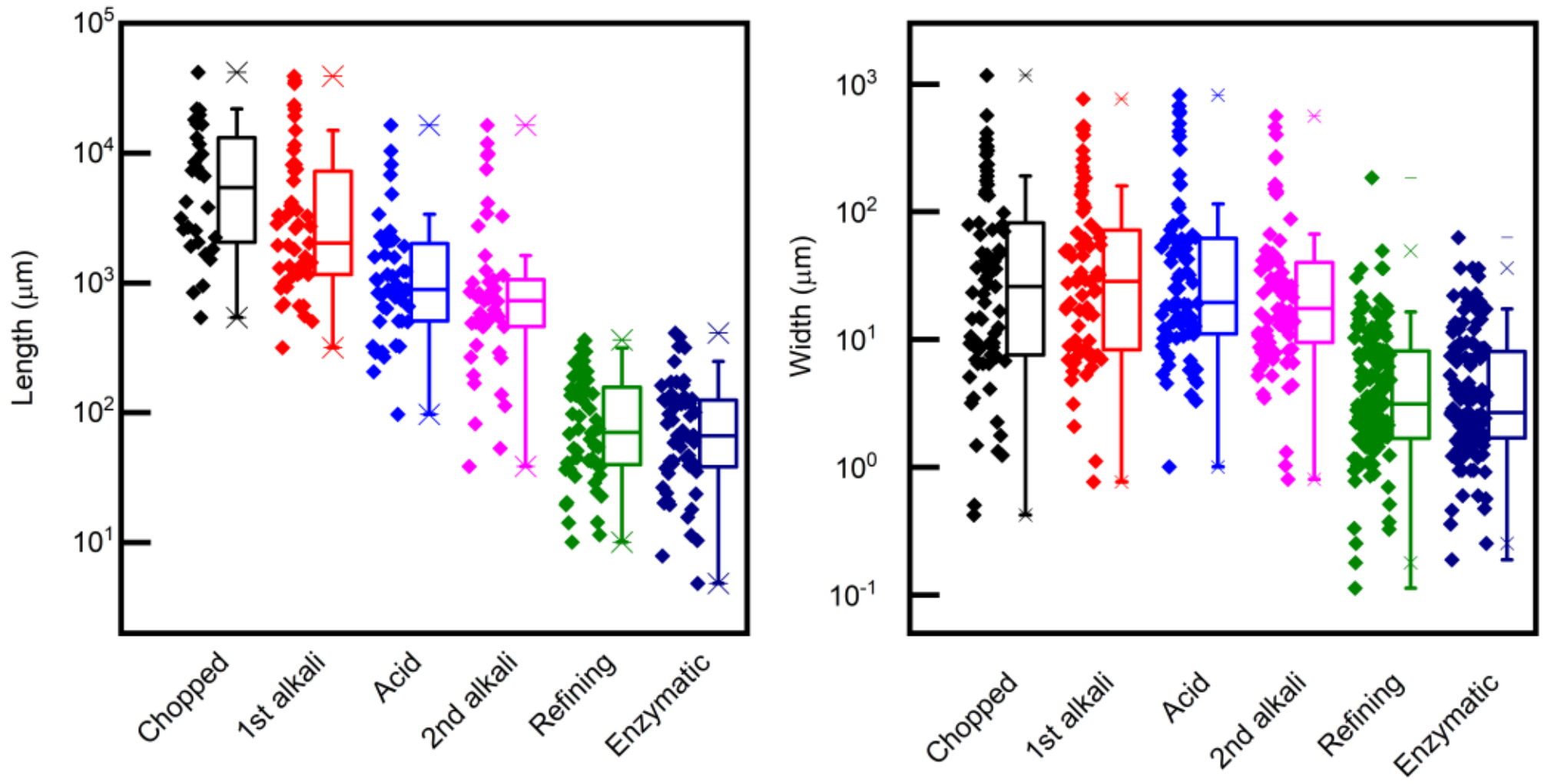

Figure 5. Boxplots of the lengths and widths of the hemp fibers after each pre-treatment step. The measured data points are shown close to the corresponding boxes.

### *3.2 Morphology and dimensions of the nanocellulose*

After the series of pretreatments, the fibers, still of micrometric dimensions, were subjected to mechanical defibrillation in an ultra-fine friction grinder, to obtain nanocellulose. The FESEM observation of the nanocellulose suspension highlighted that nanosized fibrils were present together with larger micron sized fibers (Figure 6). The nanosized fraction, separated by decanting as mentioned in section 2.4 for the inspection by TEM (Figure 7), was composed by individual rod-like fibrils with lengths of 100 – 300 nm and widths of 5 – 12 nm, and stacks of fibrils with lengths of 380 nm to 3 μm and widths between 20 nm and 200 nm. The dimensions of the individual rod-like fibrils were comparable to those of the cellulose nanocrystals (CNC) obtained by Luzi et al. from Carmagnola hemp bast fibers by acid hydrolysis and to those obtained by Xu et al. from hemp and flax bast fibers by endoglucanase treatment (Xu et al. 2013; Luzi et al. 2014).

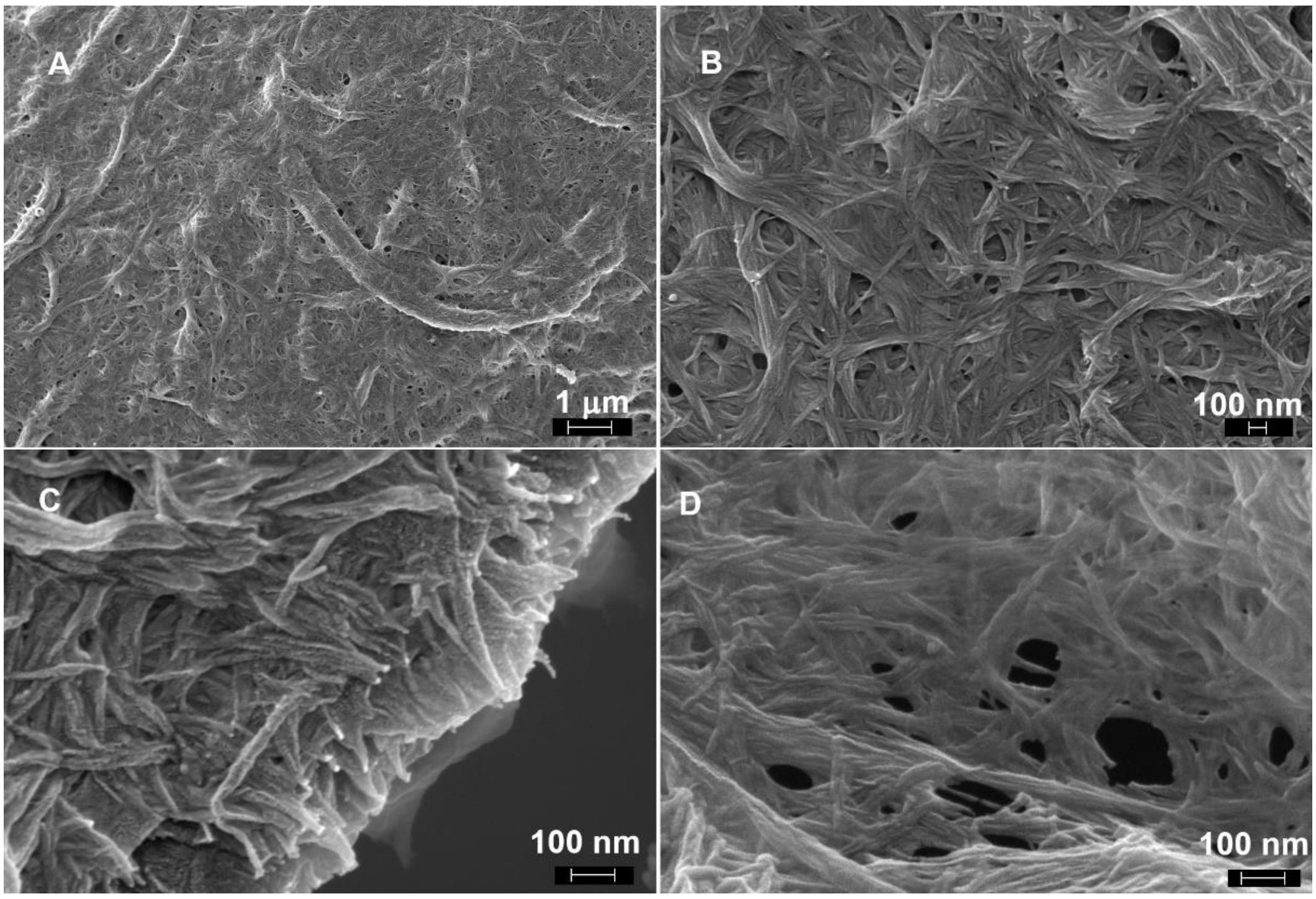

Figure 6. FESEM images of the nanocellulose, composed of nanosized rod-like fibrils mixed with micron sized fibrils

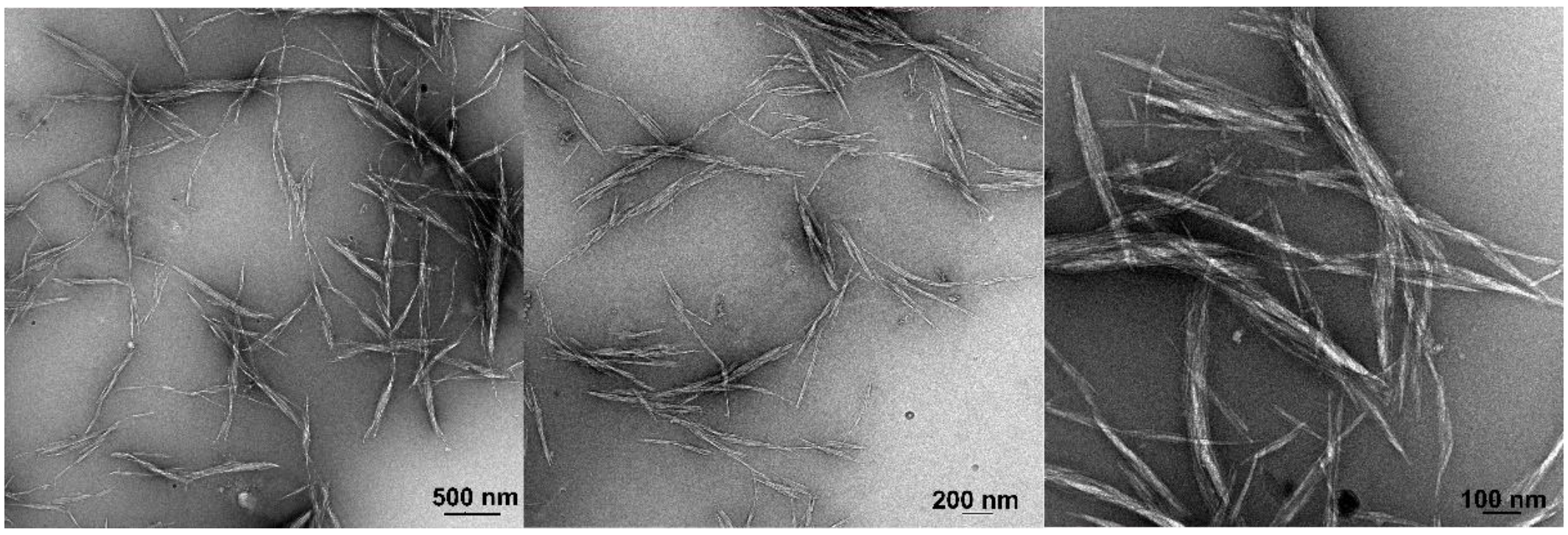

Figure 7. TEM images of the individual nanosized fibrils and of stacks of fibrils.

### *3.3 FTIR characterization*

The FTIR spectra of the raw hemp fibers and of the fibers after each treatment are shown in Figure 8. In the spectrum of the raw fibers, absorption bands that correspond to the stretching and bending vibrations of main compounds present in lignocellulosic fibers, such as cellulose, hemicellulose and lignin were identified. The peak assignments are reported in Table 1.

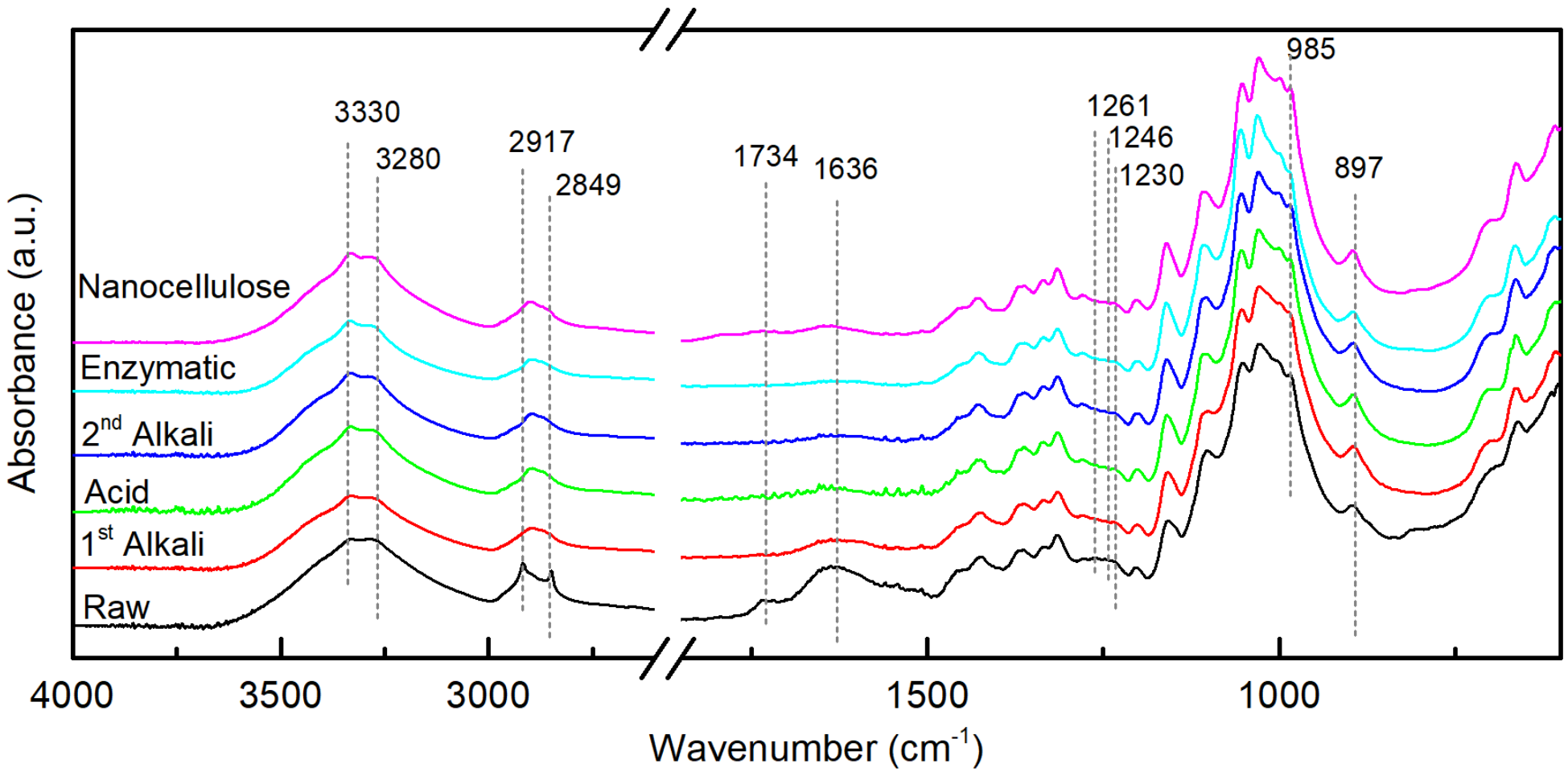

Figure 8. FTIR spectra of the hemp fibers, raw and after the pretreatments, and of the hemp nanocellulose.

Table 1 Main FTIR vibrations in cellulose, hemicellulose and lignin

| Wavenumber ($cm^{-1}$) | Vibration |
|---|---|
| 3700-3000 | O-H stretching of hydroxyl groups in polysaccharides (cellulose and hemicellulose) (Cintrón and Hinchliffe, 2015) |
| 3330 | Valence vibration of hydrogen bonded O–H (Fackler et al., 2010) |
| 3280 | O6-H6···O3 intermolecular hydrogen bond in cellulose Iβ (Boukir et al., 2019; Popescu et al., 2016; Schwanninger et al., 2004) |
| 2917, 2849 | C-H stretching of aliphatic methylene groups of fatty acids (Boeriu et al., 2004; Mondragon et al., 2014) |
| 2895 | C-H symmetrical stretching of hydrocarbon constituents in polysaccharides (Stevulova et al., 2014) |
| 1734 | C=O unconjugated stretching of acetyl groups in hemicelluloses (Abraham et al., 2016) |
| 1636 | O-H bending/deformation in water bound to cellulose (Abraham et al., 2016) |
| 1595, 1505 | skeletal vibrations in phenolic ring in lignin (Boukir et al., 2019; Fackler et al., 2010) |
| 1462 | $\delta CH_2$ asymmetric bending in cellulose I, $\delta CH_3$ asymmetric bending in lignin and hemicelluloses (Boukir et al., 2019) |
| 1424, 1420 | $\delta CH_2$ symmetric bending in crystalline cellulose I (strong), in amorphous cellulose (weak) (Boukir et al., 2019; Stevulova et al., 2014) |
| 1370 | δC-H and $\delta sCH_3$ in cellulose and hemicelluloses (Boukir et al., 2019) |
| 1337 | OH in plane bending in amorphous cellulose (Stevulova et al., 2014) |
| 1320 | $CH_2$ wagging in crystalline (Boukir et al., 2019; Stevulova et al., 2014) |
| 1261 | CO stretching, guaiacyl rings in lignin (Fackler et al., 2010; Lehto et al., 2018; Tejado et al., 2007) |
| 1246 | stretching vibration of C=O in xylan (Jiang et al., 2015) |
| 1230 | CO stretching of syringyl rings in lignin (Lehto et al., 2018) |
| 1104 | C-O vibrations from cellulose secondary alcohols (Cintrón and Hinchliffe, 2015) |
| 1051 | C–O stretching in hemicellulose, lignin and cellulose secondary alcohols (Abraham et al., 2016) |
| 1034 | C–O–C skeletal vibration of polysaccharides ring (Boukir et al., 2019) |
| 1027 | C–C, C–OH, C–H ring and side group vibrations in hemicellulose, pectin and cellulose primary alcohols (Stevulova et al., 2014) |
| 1002 | C-O vibrations from cellulose primary alcohols (Cintrón and Hinchliffe, 2015) |
| 985 | C-O bending of cellulose alcohols involved in hydrogen bonds (Cintrón and Hinchliffe, 2015) |
| 897 | glycosidic bonds symmetric ring-stretching mode of amorphous regions in cellulose (Stevulova et al., 2014) |

The first alkali pretreatment caused the disappearance of the two sharp peaks at 2917 $cm^{-1}$ and 2849 $cm^{-1}$, indicating the removal of fatty acids. Furthermore, the peaks at 1734 $cm^{-1}$ and 1246 $cm^{-1}$, as well as at 1261 $cm^{-1}$ and 1230 $cm^{-1}$ decreased, confirming that the content of hemicellulose and lignin, respectively, diminished. The peak at 1636 $cm^{-1}$ assigned to bound water strongly decreased. Peaks at ca. 1104 $cm^{-1}$, 1053 $cm^{-1}$ and 1029 $cm^{-1}$, that can be assigned to cellulose alcohols, became sharper. The broad band between 3700 $cm^{-1}$ and 3000 $cm^{-1}$ showed some changes, particularly in the intensities at 3330 $cm^{-1}$ and 3280 $cm^{-1}$, that may indicate changes in the type of hydrogen bonds formed within and between the cellulose fibrils.

The following acid and second alkali pretreatments caused a further slight change of the spectra following the same trends, indicating the removal of some residual non-cellulosic compounds, in agreement with the visual observation reported in section 3.1. After the enzymatic pretreatment and after the defibrillation in the grinder, only some changes in the relative heights of the peaks in the region from 1104 $cm^{-1}$ to 985 $cm^{-1}$ were detected, possibly indicating changes in the conformation of the hydroxyl groups of cellulose. The enzyme pre-treated fibers showed a lower intensity of the peak at 3280 $cm^{-1}$ in comparison to that of the peak at 3330 $cm^{-1}$, suggesting the disruption of intramolecular hydrogen bonds, as well as a lower intensity of the bands at 1002 $cm^{-1}$ and 985 $cm^{-1}$, compared to the neighboring peaks at 1056 $cm^{-1}$ and 1030 $cm^{-1}$. The spectrum of the dry nanocellulose on the contrary showed an increased intensity of the peaks at 3280 $cm^{-1}$ and at 1002 $cm^{-1}$ and 985 $cm^{-1}$; the increase of the peak at 3280 $cm^{-1}$ suggests that more intermolecular hydrogen bonds were formed upon drying.

*3.4 X-ray diffraction*

Native cellulose typically consists of two crystalline forms, Iα and Iβ that can coexist in a single microfibril in various proportions depending on the cellulose source. Cellulose Iα, which is predominant in e.g. algae or bacterial cellulose, has a triclinic unit cell, containing one cellulose chain. Cellulose Iβ, which is the main component in fibers from higher plants such as hemp, cotton and the like, has a monoclinic unit cell, containing two parallel cellulose chains (French, 2014; Hult et al., 2003; O'Sullivan, 1997). All fiber samples in Figure 9 clearly showed diffraction peaks at 2θ values of 14.8°, 16.5°, 22.5°, and 34.5°, corresponding to the ($1\bar{1}0$), (110), (200), and (004) crystallographic planes of semi-crystalline cellulose Iβ. Only the raw hemp showed an additional peak at a 2θ of 25.9°, which could not be attributed to any of the constituents of natural fibers. As the peak disappeared after rinsing the fibers in cold distilled water, it was attributed to the presence of impurities. Cellulose II exhibits characteristic peaks for the ($1\bar{1}0$), (110) and (020) reflections located at 12.2°, 20° and 22.1°, respectively. The shoulder around 20° that appeared in the diffractograms of all fiber samples after the first alkali pretreatment, may correspond to the superposition of a weak (110) reflection of cellulose II and of (012) and (102) reflections of cellulose Iβ appearing in randomly oriented samples (Fawcett et al., 2013; French, 2014). A very weak peak at 12.2° was only visible in the nanocellulose pattern. Strong alkali treatments are known to induce a partial transformation of cellulose I to the more stable cellulose II structure, the so-called mercerization.

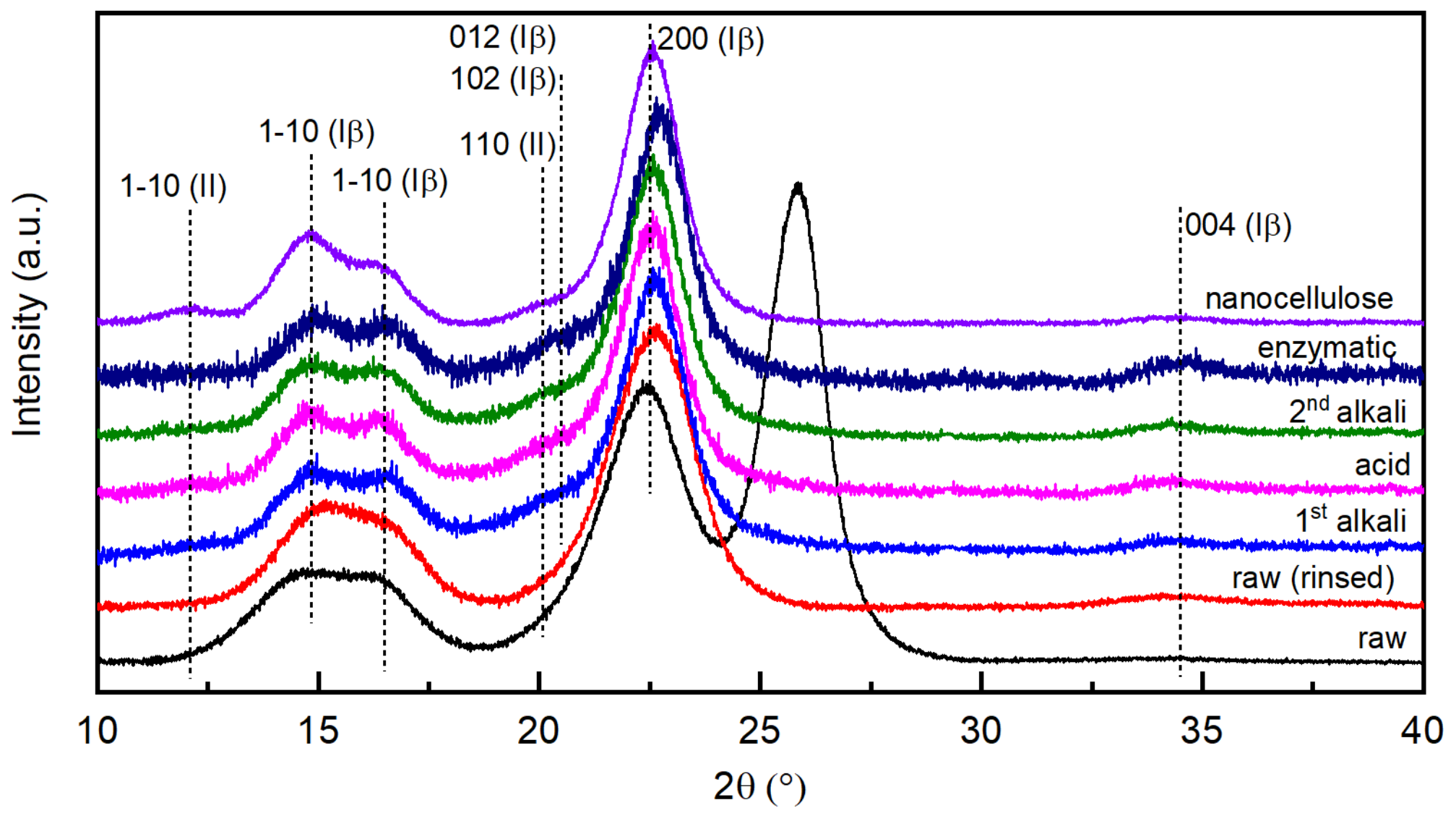

Figure 9. X-ray diffractograms of the hemp fibers at different stages of the process

This happens particularly in combination with bleaching treatments (Abraham et al., 2016; Dinand et al., 2002; Jiao and Xiong, 2014; Luzi et al., 2014). On the other hand, the presence of residual hemicellulose and lignin, which in our case is suggested by FTIR (section 3.3) and thermogravimetric analysis results (section 3.5), is thought to inhibit the transformation, possibly by hindering the penetration of alkali in the cellulose fibrils (Luzi et al., 2014). This is consistent with finding only a negligible presence of cellulose II in the samples after the alkali treatments.

To estimate the size of the crystallites we have assumed that the lateral crystallite size is the primary contributor to broadening of X-ray peaks. The crystallite size was determined from the broadening of the (200) reflection because it was the most clearly resolved of the three X-ray reflections. The dimension which is measured is in a direction normal to the surface of a crystallite. Table 2 summarizes the positions of the fitted peak and the width of the peak at half the height (β), the degree of crystallinity (CI) and the size of the crystallites.

**Table 2** Values of $2\theta_{max}$ (position of the fitted peak) and β (width of the peak at half the height) for the 200 reflection, degree of crystallinity (CI), and size of the crystallites determined from application of the Scherrer equation to 200 reflection for the hemp fibers, raw and after the pre-treatment step indicated in the first column

| Fiber sample | 200 (±0.05°) | | CI (%) | Crystallite size (Å) |
|---|---|---|---|---|
| | $2\theta_{max}$ | β | | |
| Raw hemp | 22.41 | 2.18 | 74 | 39 |
| 1st Alkali | 22.71 | 1.62 | 79 | 55 |
| Acid | 22.57 | 1.60 | 80 | 55 |
| 2nd Alkali | 22.58 | 1.47 | 82 | 60 |
| Enzymatic | 22.66 | 1.45 | 80 | 61 |
| Nanocellulose | 22.58 | 1.45 | 80 | 61 |

The degree of crystallinity increased abruptly after the first alkali treatment, most probably due to the removal of amorphous materials such as lignin, hemicellulose, waxes. At the same time the crystallite size increased as well. Then crystallinity slightly increased up to 80 % for the final nanocellulose. The progressive decrease of the width at half height values (β) following the chemical pretreatments is reflected in the increase of crystallite sizes. A correlation between crystallite sizes and CI was found, in agreement with other researchers that observed an inverse correlation between X-ray broadening (β) and X-ray fractional crystallinity (Gjönnes et al., 1958; Gjønnes et al., 1958; Krässig, 1993). This is consistent with recent theory that amorphous cellulose concentrates at the surface of crystallites (Garvey et al., 2005). Smaller crystallites have a larger surface area to volume ratio, therefore a larger fraction of amorphous cellulose surrounding the crystalline part, leading to lower CI, compared to larger crystallites, which have instead smaller surface area to volume ratios, therefore lower amounts of amorphous cellulose surrounding the crystalline fraction, resulting in higher CI.

*3.5 Thermogravimetric analysis*

The results of the thermogravimetric analysis performed on the raw and pretreated hemp fibers, and on the nanocellulose are reported in Figure 10.

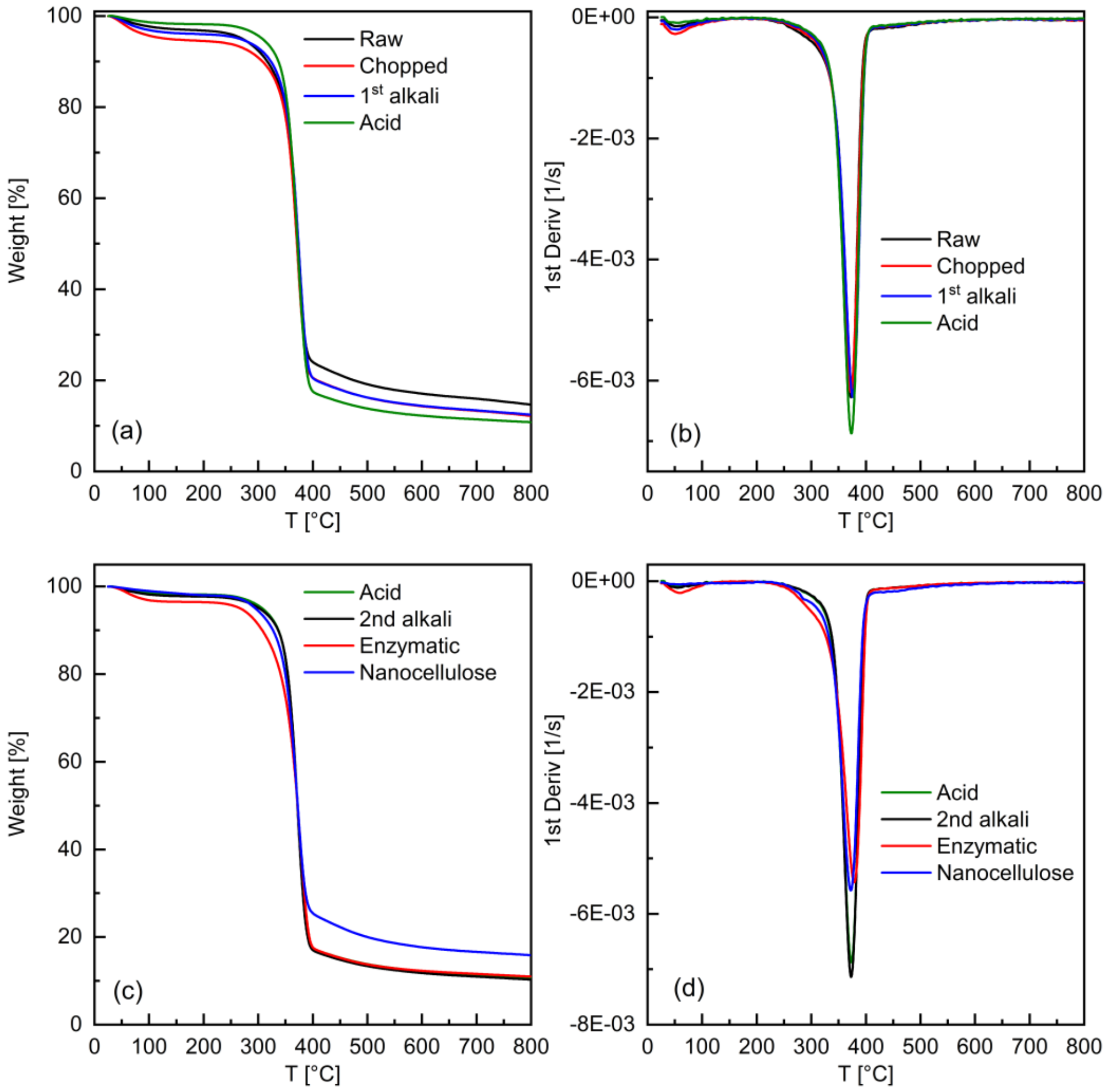

Figure 10. Thermogravimetric analysis under $N_2$ of the fibers, as received and after different pretreatment steps: (a,c) evolution of the weight as a function of temperature, and (b,d) first derivative of the weight curve.

Generally, all the thermogravimetric curves presented the following features: i) an initial weight loss due to evaporation of residual water, around 100 °C; ii) in the 350-400°C region the degradation of cellulose; iii) between 250 and 300 °C the degradation of hemicellulose; iv) from 250 to 600 °C the degradation of lignin, that usually happens in a wide range of temperatures, due to the complexity of the lignin structure; v) a residue of 11-17 wt% .

Comparing in details the thermal behavior of the raw and chopped hemp fibers with that of fibers treated with 10 wt% NaOH, one notices only a very slight decrease of the two shoulders near the main peak and a slight increase of the temperature of maximum degradation rate for cellulose. These small changes can be attributed to the removal of non-cellulosic compounds at the surface of the fibers. After the acid pretreatment, with 1M HCl, the residual water present in the fibers, associated with the low temperature weight loss, decreased to less than 2%, consistently with the decrease of the associated peak in FTIR analysis. The maximum rate for cellulose degradation increased, while the residual weight decreased to about 11%. These effects are attributed mostly to acid hydrolysis of hemicellulose and cellulose, leading to removal of hemicellulose and molecular weight decrease of cellulose (Oriez et al., 2019). The second alkali pretreatment, consistently to what seen by FTIR analysis, caused only minor changes in the same direction as the acid pretreatment. After the enzymatic pretreatment, the amount of residual water increased and the maximum degradation rate decreased: finer fibers are likely to form quite densely packed structures, hindering both water removal during the pre-drying and thermodegradation. The onset of cellulose thermodegradation was at lower temperature, probably due to the reduction of the molecular weight caused by the endoglucanase action.

Main differences of the dry nanocellulose with respect to the enzymatically treated sample are an initial slow and almost linear weight loss (2-3%) until about 190 °C, which can be attributed to residual non-freezing water, and the increase of both the onset temperature for cellulose degradation and of the amount of the residue, due to the formation of a denser structure upon drying.

The DSC thermogram of the dry nanocellulose (Figure 11) during the first heating run showed a broad endotherm peak between 0 °C and 200 °C, centered around 90 – 95 °C, which was not present in the second heating run. The presence of this peak, attributed to the evaporation of non-freezing water, is consistent with what seen in thermogravimetric analysis. No other transitions appeared in the observed temperature range.

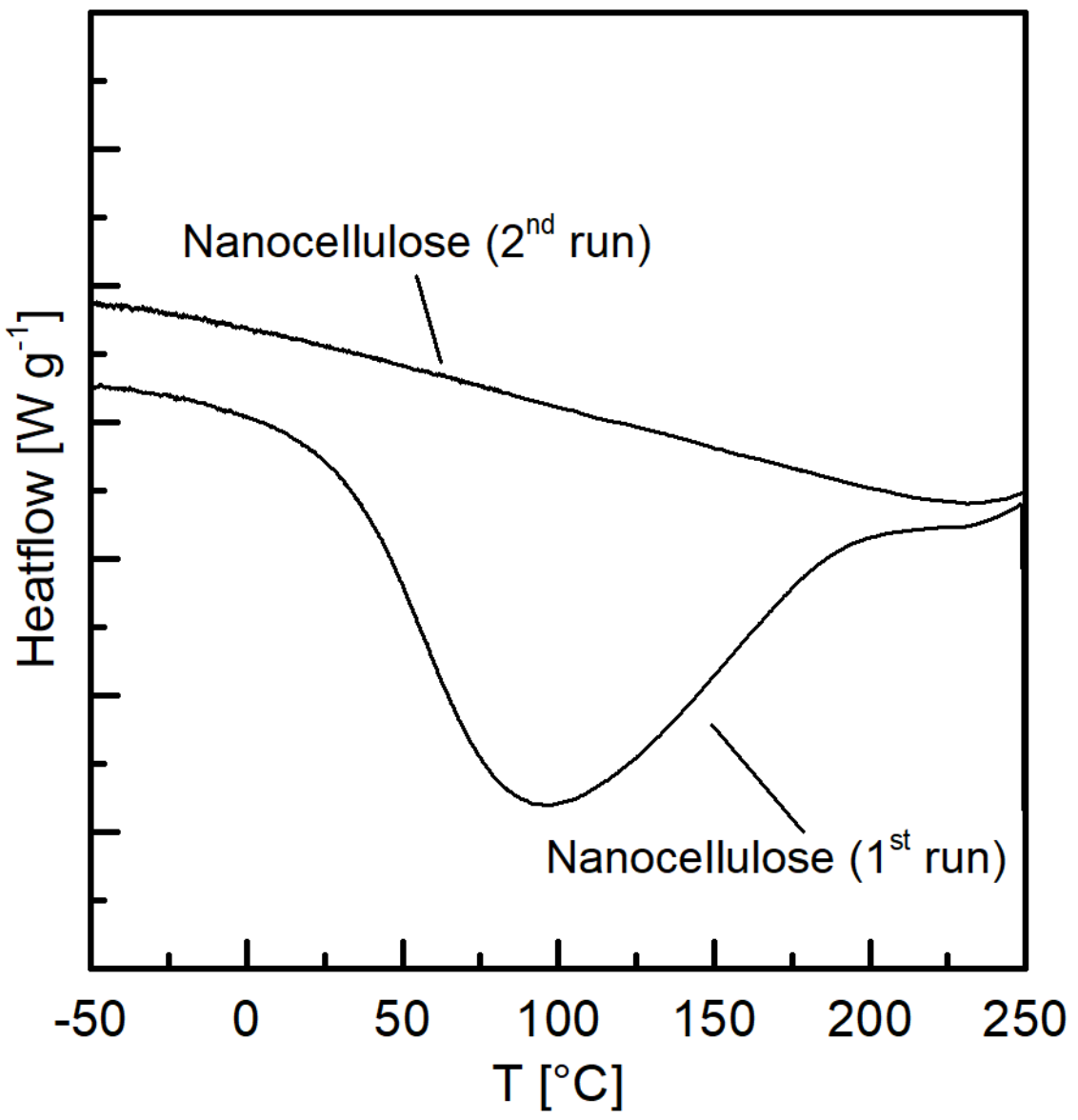

Figure 11. DSC thermogram of dry nanocellulose

*3.6 Compatibility of nanocellulose with solvents*

As the nanocellulose is an efficient reinforcing agent in polymeric composites, where usually the matrix has a low polarity, their compatibility with different solvents was tested. Samples of the suspension collected after the ultrafine friction grinding were diluted by adding either distilled water, ethanol or acetone, to obtain a solid content of 0.25 wt%. The obtained suspensions were left to stand at room temperature for 12 days, as shown in Figure 12. The suspension in water was stable, showing no settling. In water/ethanol and water/acetone there was some settling of the cellulose fibrils; while after 1 hour only the water/acetone solution showed some settling, after 12 days a large settling was visible both in water/ethanol and in water/acetone. The content of ethanol and acetone in the solutions was close to 82 wt%, and the density of the solutions was estimated to be close to 840 kg/m$^3$ (Khattab et al., 2012; Noda et al., 1982). The settling in the organic solvent mixtures may be attributed to the lower density with respect to water, which allows a higher settling velocity; destabilization of the fibrils in the presence of the organic solvents may also be ascribed to their different polar/dispersive character.

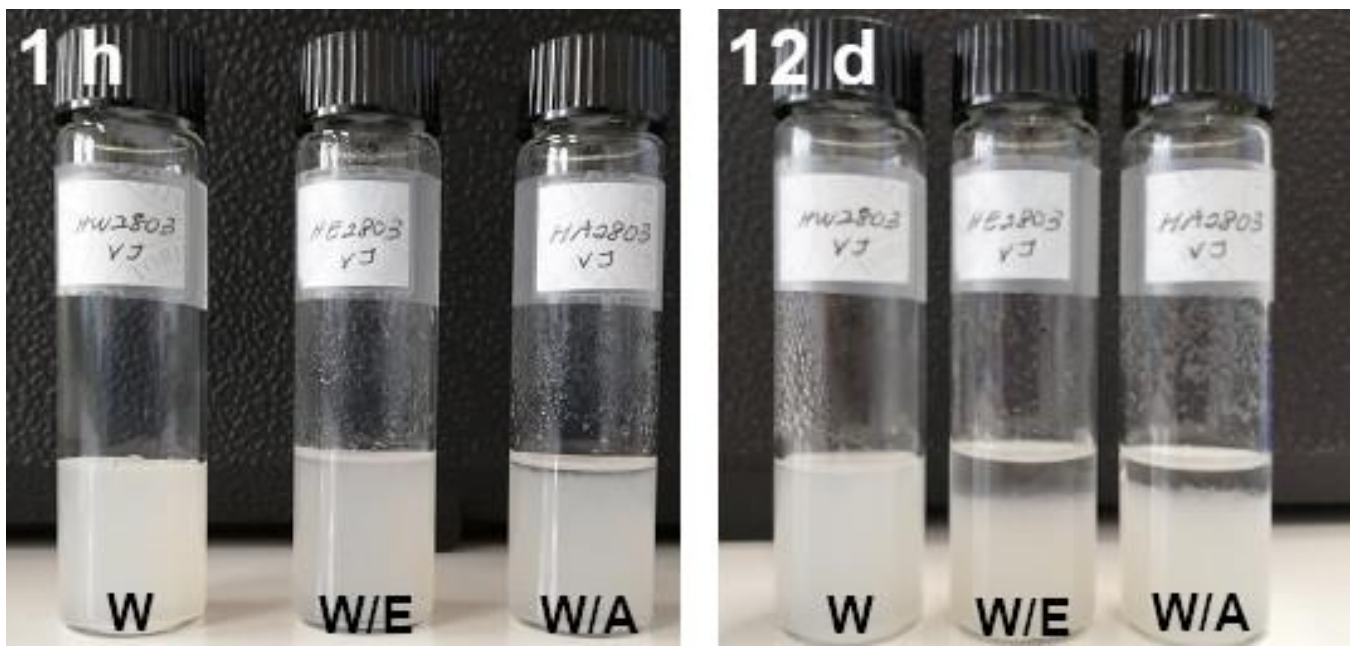

Figure 12. Hemp nanocellulose after settling for 1 hour (1h) and 12 days (12 d) in different solvents: W = water, W/E = water/ethanol, W/A = water/acetone)

*3.7 Use of nanocellulose for the preparation of nanopapers and their characterization*

The nanocellulose was used to prepare nanopapers by a standard procedure based on filtering the suspensions through a Rapid-Köthen standard sheet former and drying. The handsheet obtained from the hemp nanocellulose was transparent with a brownish color, and relatively flexible. Photos of samples cut from the handsheet are shown in  Figure 13.

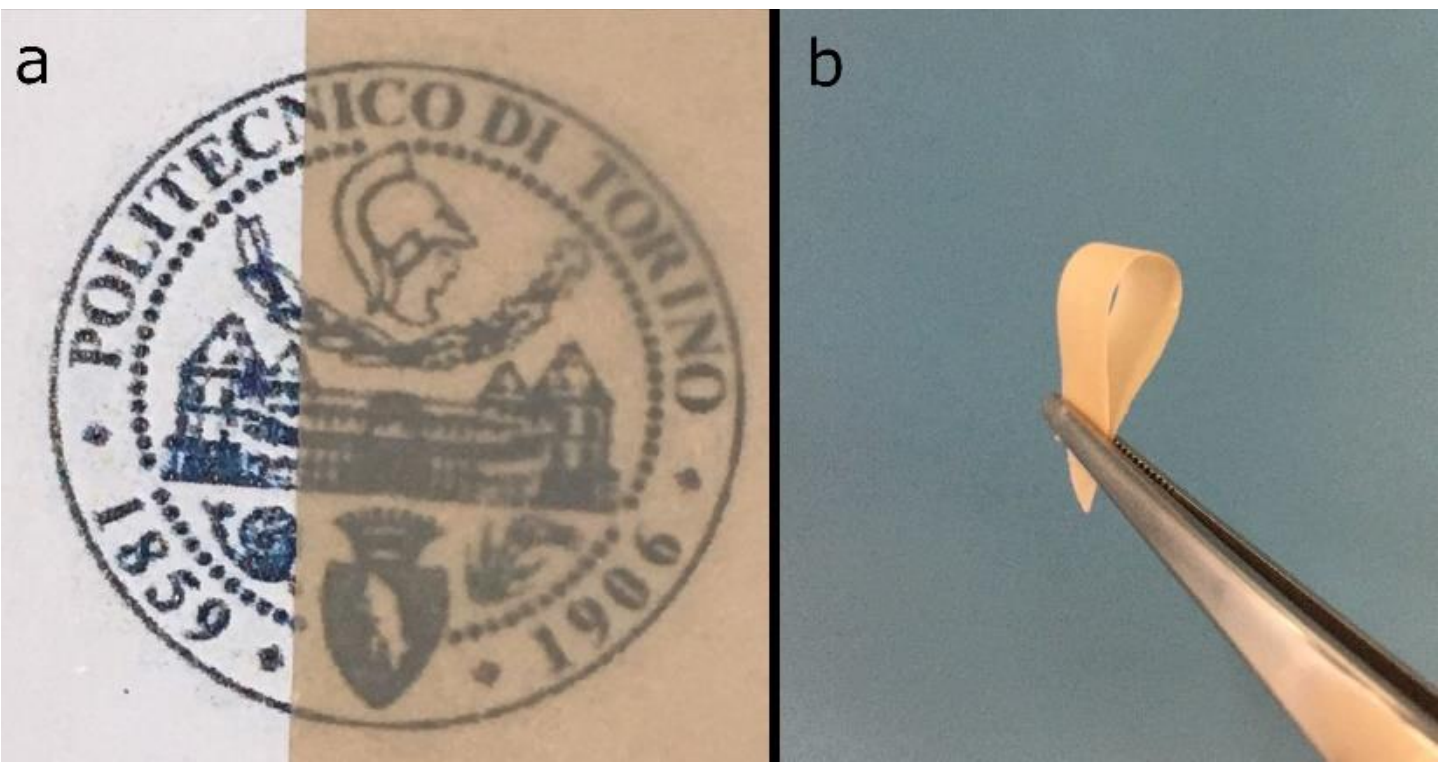

Figure 13. Photos of samples of the hemp MFC handsheet showing (a) color and transparency, and (b) flexibility

The optical microscopy observation of the surface revealed the presence of a larger scale structure (Figure 14a) repeating the pattern of the PA 6.6 fabric onto which the handsheet was prepared (Figure 14c). At a higher magnification, the microstructure of the cellulose fibrils appeared (Figure 14b). Observation with a confocal microscope allowed a 3D reconstruction of the sample surface (see Figure S1).

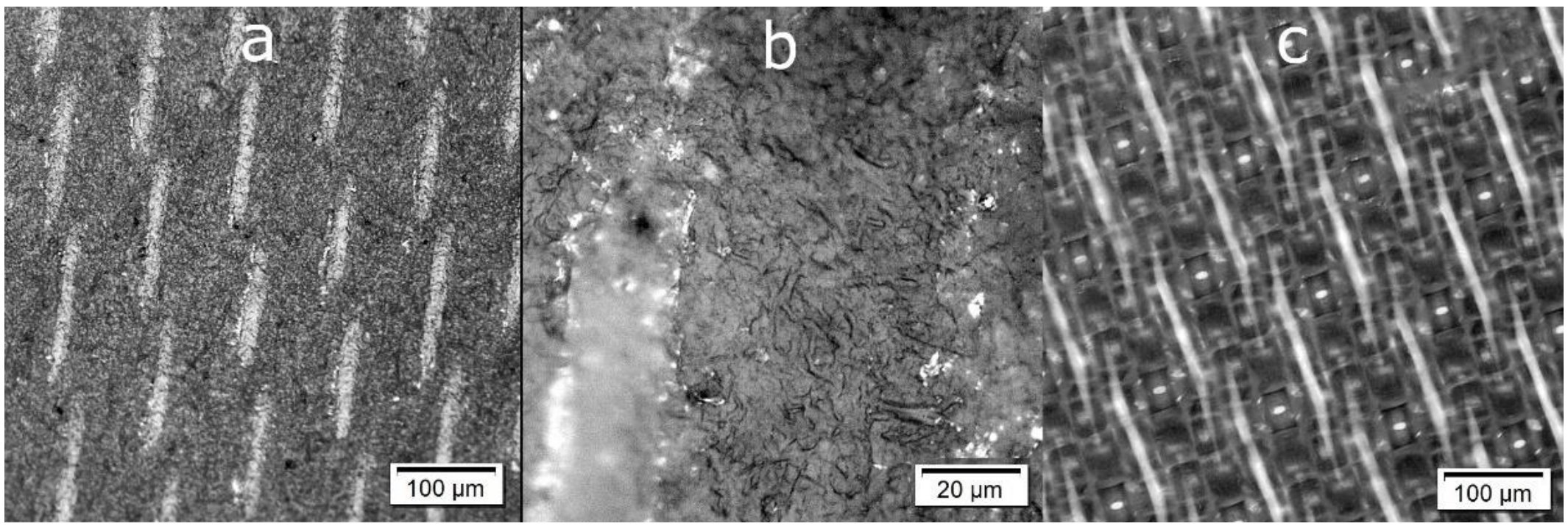

Figure 14. Optical micrographs of (a,b) the surface of the handsheet and (c) the surface of the polyamide 6.6 fabric used as filter in the hansdheet formation process.

The results of the DSC and TGA analysis of the handsheet did not show any relevant difference with those reported for the dry nanocellulose (see Figures S2 and S3).

The dynamic mechanical analysis performed on strips cut from the handsheet (Figure S4) showed a storage modulus close to 5 GPa between -100 °C and 180 °C, and did not highlight any phase transition in the entire temperature range. The tensile tests (Figure S5) revealed an elastic behavior, and the rupture was brittle. The Young's modulus was of the order of 9 ± 2 GPa in agreement with the results reported for films obtained from cellulose nanocrystals ranging from 3 to 30 GPa (Meng and Wang, 2019). Elongation at break was in the 0.15 - 0.35% range, and tensile strength was 20 ± 5 MPa, which is lower than literature data ranging from 30 to 130 MPa. The difference can be ascribed to defects of the specimens, in particular small flaws appearing on the sides upon cutting and visible at the naked eye: in fact, the material was very brittle.

Finally, in view of assessing the potential of this material for packaging applications, a preliminary assessment of the barrier properties of the 55 µm thick handsheet, was performed. The water vapor transmission rate WVTR at 38 °C and 50% RH was 56 ± 8 (i.e. $P_{WV}$ = 94 ± 13). OTR at 23 °C and 50% RH was 3 ± 2 (i.e. $P_{O2}$ = 0.2 ± 0.1). These results are in line with those reported by (Kumar et al., 2014) for nanopapers produced with rod-like nanofibrils from wood sources, having OTR values between 162 and 218, and WVTR values between 2.6 and 16.1, for a thickness of 25 µm.

## 4. Conclusion

We proved that nanocellulose can be obtained from scrap bast fibers, byproduct of hemp crops harvested for seed, separated from the shives with a disordered mechanical method. The process included mechanical and chemical treatments, but avoided bleaching or oxidation. Thus, it is suggested that these steps, often involving environmentally hazardous chemicals, may not be necessary, unless the color of the fibers has specific application requirements.

It was possible to use the unbleached nanocellulose to fabricate nanopapers: the handsheets were transparent, and showed a brownish color. They had thermal, mechanical and permeability properties in line with similar cellulosic materials obtained from non-waste biomass, such as fibers from wood sources.

Thus, the hemp is a good source of valuable nanomaterials: the nanocellulose obtained from is not only suitable for making nanopaper as shown here, but also for nanocomposites. These were in fact successfully prepared as described in a separate manuscript (Dalle Vacche et al., 2021b).

## Acknowledgments

The authors thank the NanoBio-ICMG Platform (FR 2607, Grenoble) for granting access to the Electron Microscopy facility.

## Funding

The project ComBIOsites has received funding from the European Union’s Horizon 2020 research and innovation programme under the Marie Skłodowska-Curie grant agreement No 789454.

## Availability of data and material

The research data generated during the current study are available in the open access Zenodo repository [dataset] (Dalle Vacche et al., 2021a)

## SUPPLEMENTARY INFORMATION

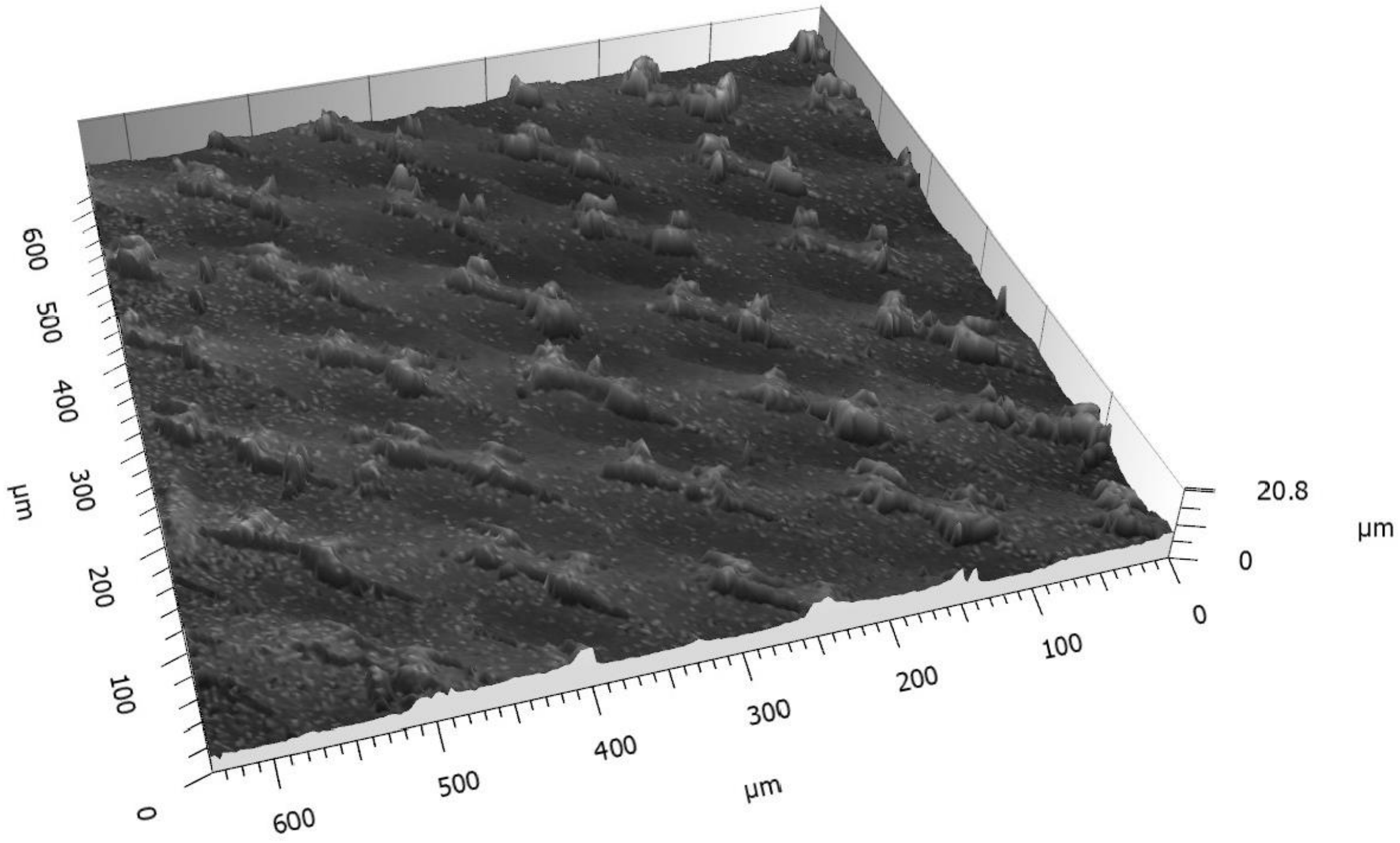

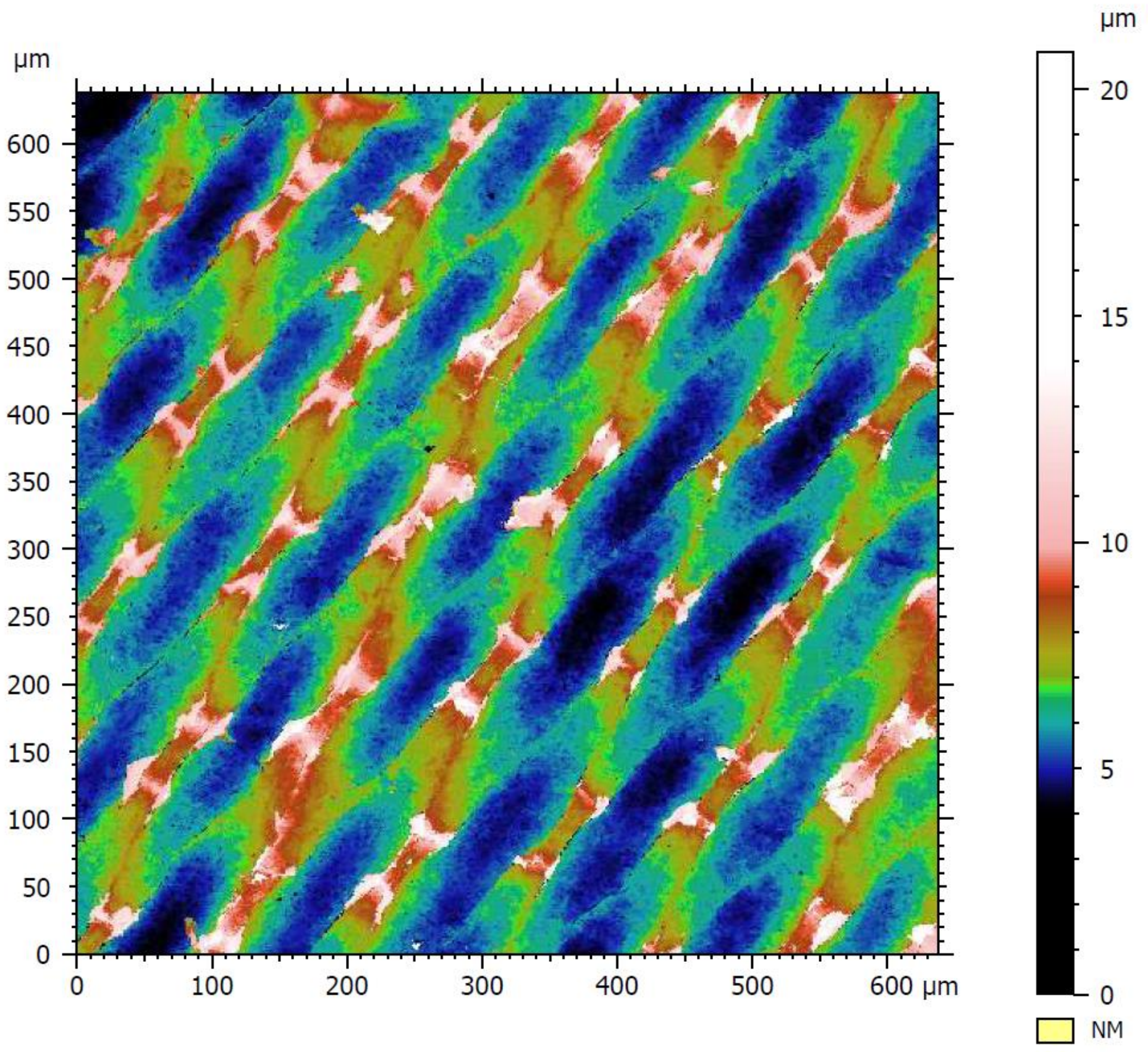

Figure S1. Images of the surfaces of the handsheet obtained with a ZEISS LSM900 confocal microscope (Carl Zeiss Microscopy GmbH, Germany)

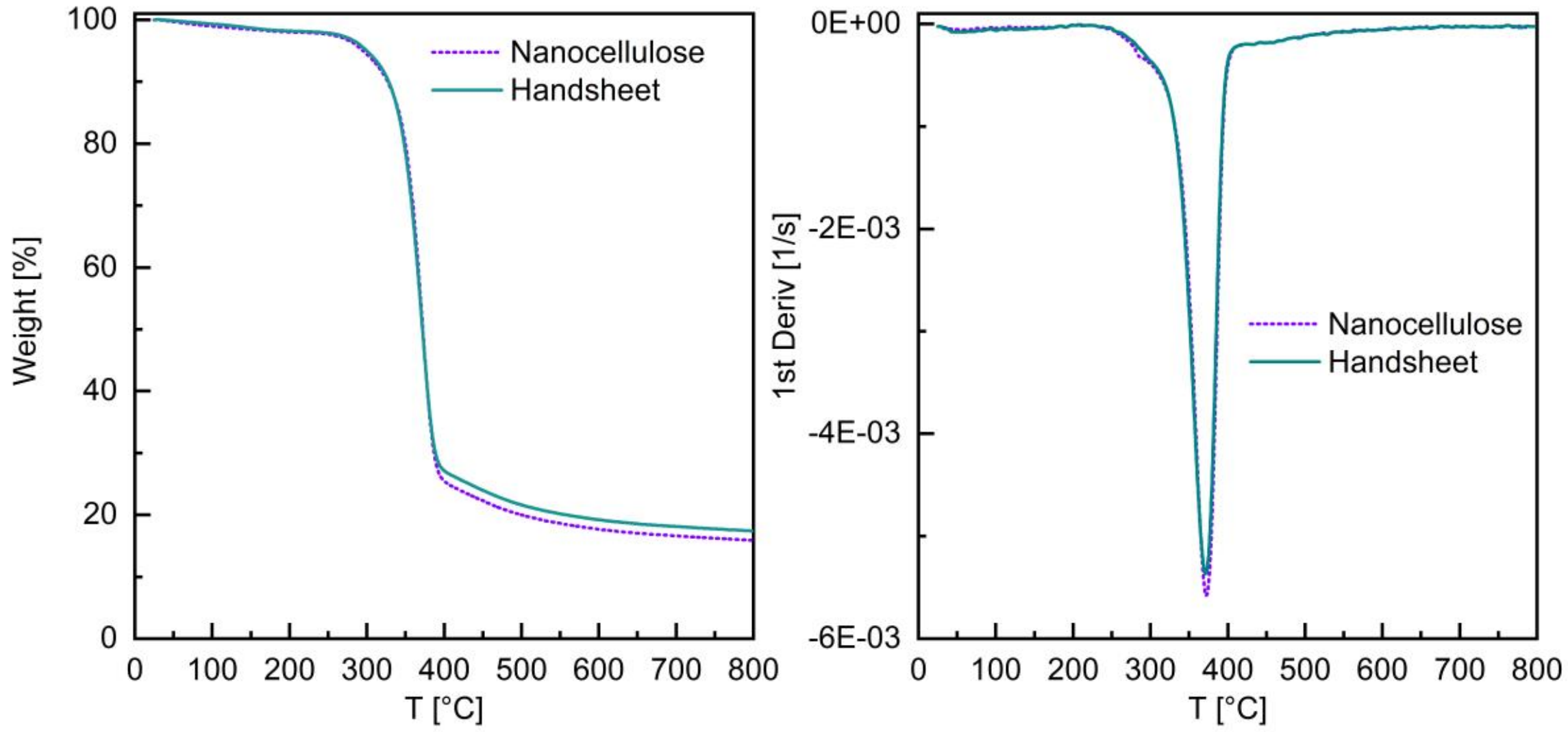

Figure S2. Comparison of the weight and first derivative curves obtained from the thermogravimetric analysis of the nanocellulose and the handsheet

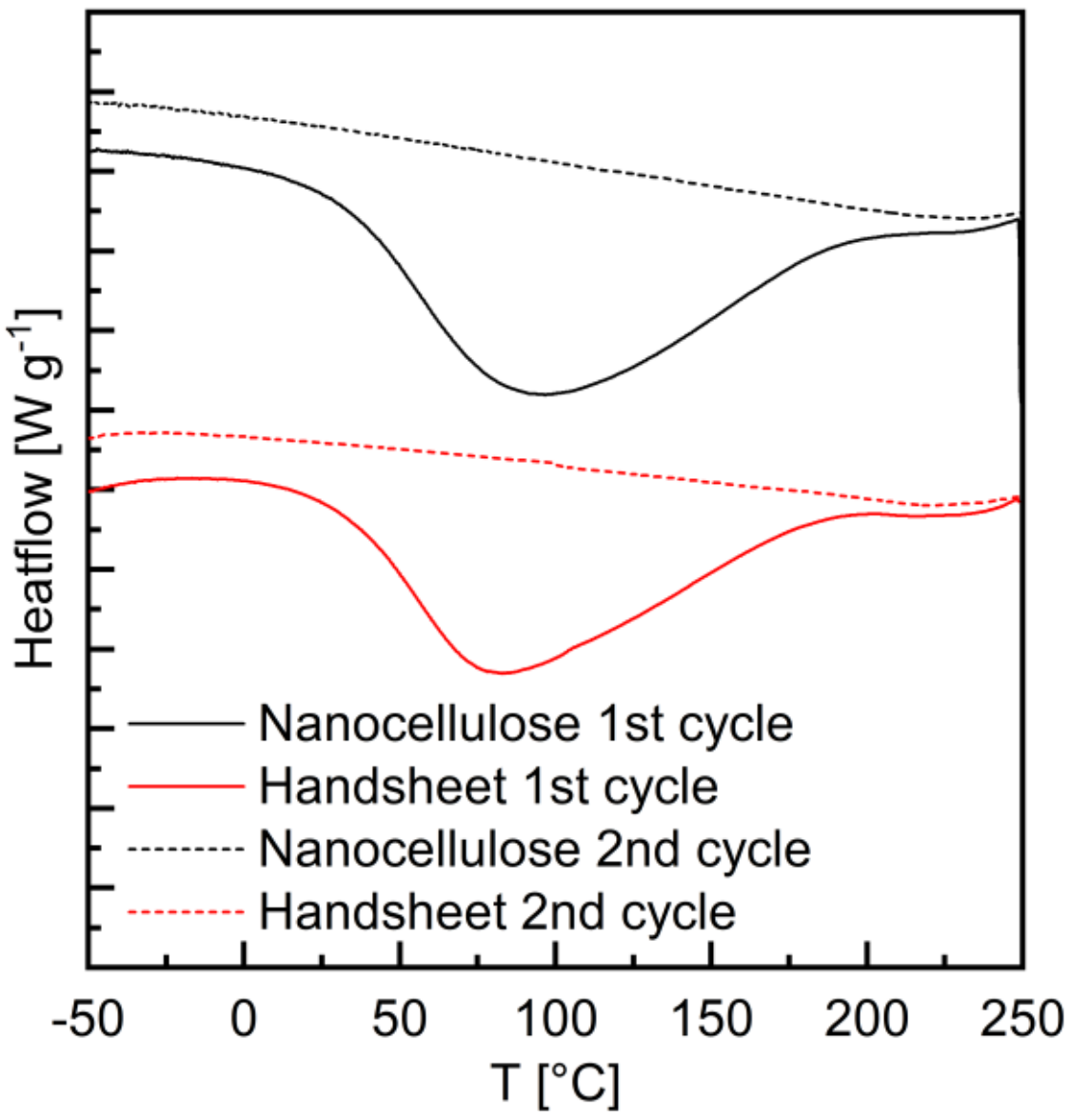

Figure S3. Comparison of the dynamic scanning calorimetry results obtained for the nanocellulose and the handsheet

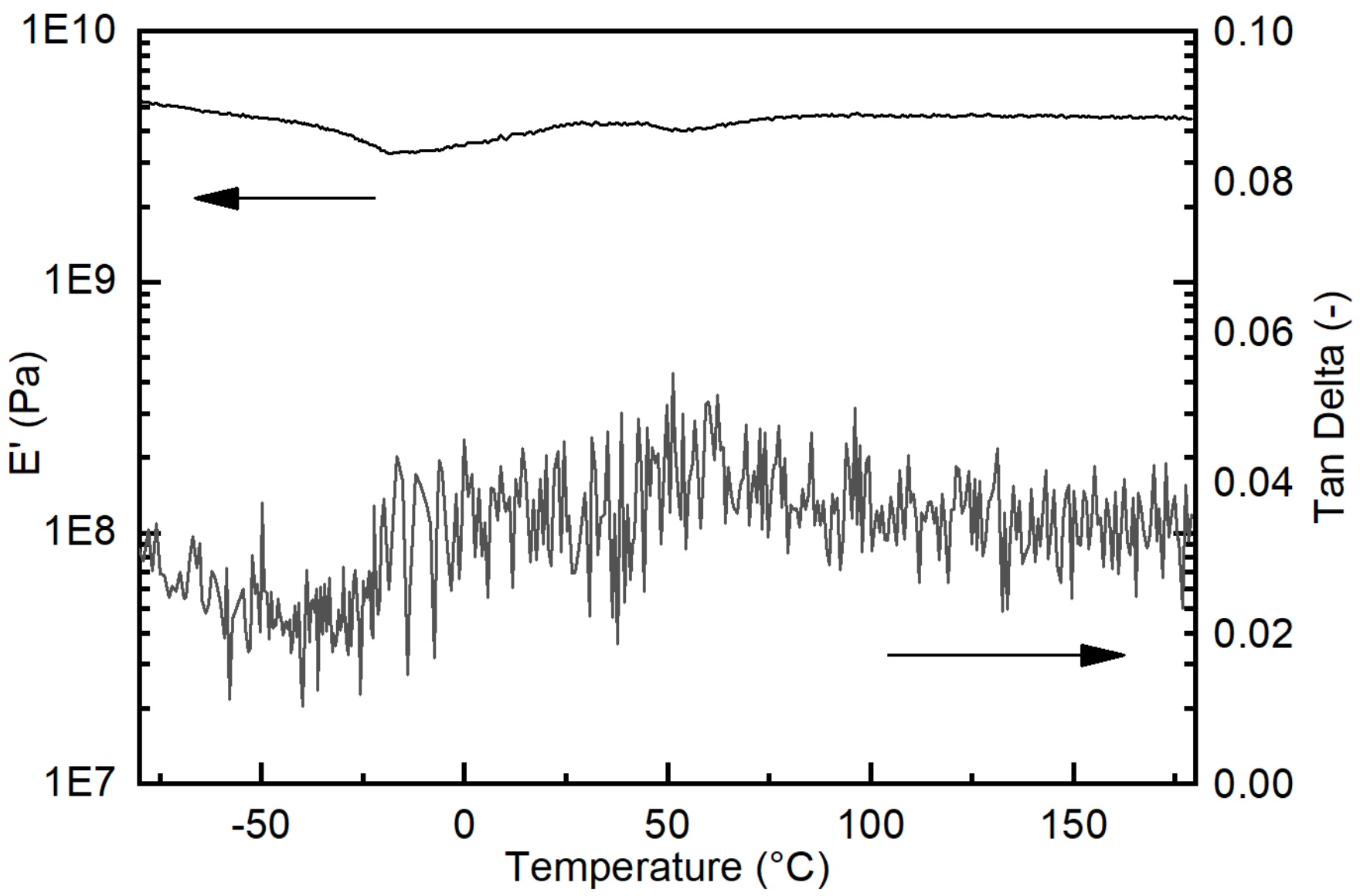

Figure S4. Results of the dynamic mechanical analysis performed on strips cut from the hemp handsheet

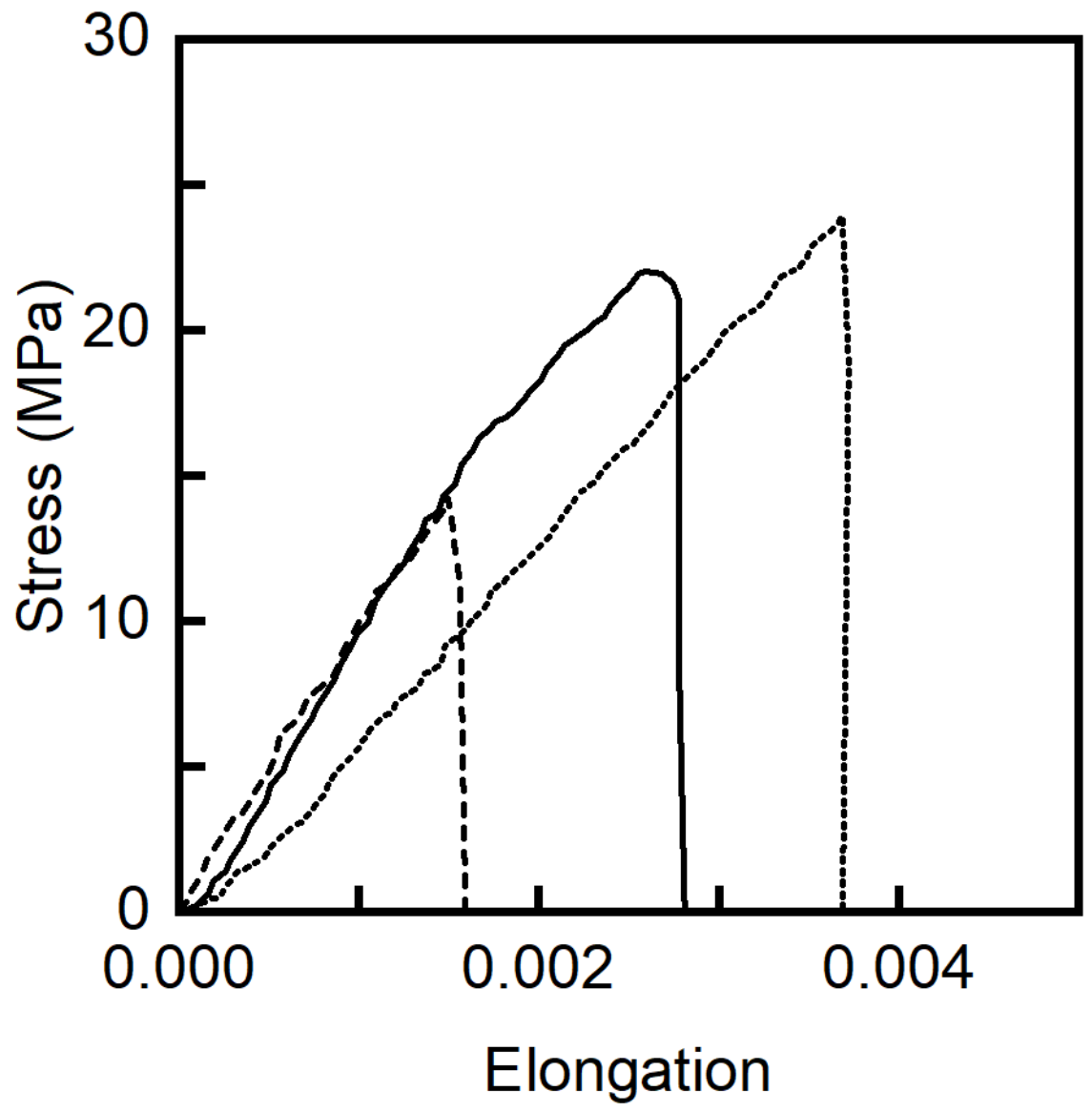

Figure S5.  Stress-strain curves obtained in the tensile test performed on strips cut from the hemp handsheet